\documentclass[journal=jctcce,manuscript=article,layout=twocolumn]{achemso}

\usepackage[colorlinks,linkcolor=blue,citecolor=blue,urlcolor=black,bookmarks=false,hypertexnames=true]{hyperref} 

\usepackage[version=3]{mhchem}
\usepackage{amssymb}
\usepackage{color}
\usepackage{braket}
\usepackage{xspace}
\usepackage{cleveref}
\usepackage{graphicx}
\usepackage{subfig}
\usepackage{threeparttable}
\usepackage{textcomp}
\usepackage{setspace}
\usepackage{caption}

\usepackage{array}
\newcolumntype{L}[1]{>{\raggedright\let\newline\\\arraybackslash\hspace{0pt}}m{#1}}
\newcolumntype{C}[1]{>{\centering\let\newline\\\arraybackslash\hspace{0pt}}m{#1}}
\newcolumntype{R}[1]{>{\raggedleft\let\newline\\\arraybackslash\hspace{0pt}}m{#1}}

\makeatletter
\let\l@addto@macro\relax
\makeatother
\usepackage[fontsize=11pt]{scrextend}

\let\oldmaketitle\maketitle
\let\maketitle\relax

\renewcommand{\d}[2]{\delta_{{#1}}^{{#2}}}

\renewcommand{\c}[1]{a^\dagger_{#1}}
\renewcommand{\a}[1]{a_{#1}}

\newcommand{\om}{\omega}
\newcommand{\si}{\sigma}

\newcommand*{\bohr}{\ensuremath{a_0}\xspace}

\newcommand*{\maxe}{$\Delta_{\mathrm{MAX}}$\xspace}
\newcommand*{\mae}{\ensuremath{\Delta_{\mathrm{MAE}}}\xspace}
\newcommand*{\std}{\ensuremath{\Delta_{\mathrm{STD}}}\xspace}

\crefname{figure}{Figure}{Figures}
\crefname{table}{Table}{Tables}
\crefname{equation}{Eq.}{Eqs.}
\crefname{section}{Section}{Sections}
\crefname{subsection}{Section}{Sections}

\author{Ilia M.\ Mazin}
\email{mazin.3@osu.edu}
\affiliation{%
     Department of Chemistry and Biochemistry,
     The Ohio State University,
     Columbus, Ohio 43210, United States
}
 \author{Alexander Yu.\ Sokolov}
 \email{sokolov.8@osu.edu}
 \affiliation{%
     Department of Chemistry and Biochemistry,
     The Ohio State University,
     Columbus, Ohio 43210, United States
 }

\begin{tocentry}
\includegraphics{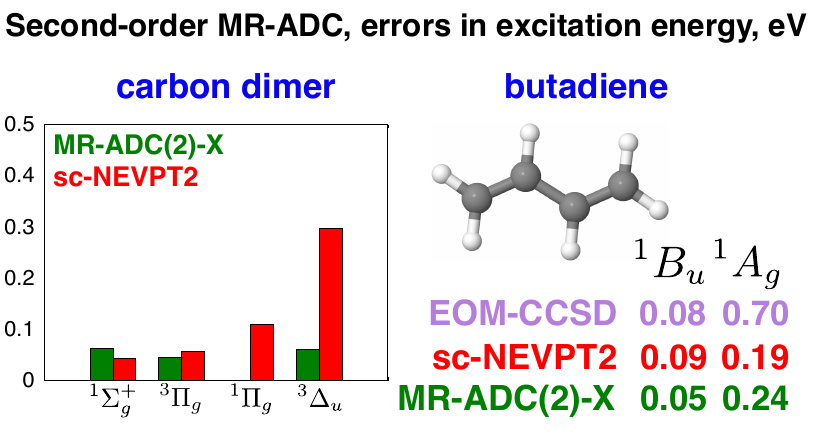}
\end{tocentry}

\title{Multireference Algebraic Diagrammatic Construction Theory for Excited States: Extended Second-Order Implementation and Benchmark}

\begin{document}

\newcommand*{\abstractext}{
We present an implementation and benchmark of new approximations in multireference algebraic diagrammatic construction theory for simulations of neutral electronic excitations and UV/Vis spectra of strongly correlated molecular systems (MR-ADC). Following our work on the first-order MR-ADC approximation [{\it J.\@ Chem.\@ Phys.\@} {\bf 2018}, {\it 149}, 204113], we report the strict and extended second-order MR-ADC methods (MR-ADC(2) and MR-ADC(2)-X) that combine the description of static and dynamic electron correlation in the ground and excited electronic states without relying on state-averaged reference wavefunctions. We present an extensive benchmark of the new MR-ADC methods for excited states in several small molecules, including the carbon dimer, ethylene, and butadiene. Our results demonstrate that for weakly-correlated electronic states the MR-ADC(2) and MR-ADC(2)-X methods outperform the third-order single-reference ADC approximation and are competitive with the results from equation-of-motion coupled cluster theory. For states with multireference character, the performance of the MR-ADC methods is similar to that of an N-electron valence perturbation theory. In contrast to conventional multireference perturbation theories, the MR-ADC methods have a number of attractive features, such as a straightforward and efficient calculation of excited-state properties and a direct access to excitations outside of the frontier (active) orbitals.
\vspace{0.25cm}
}

\twocolumn[
\begin{@twocolumnfalse}
\oldmaketitle
\vspace{-0.75cm}
\begin{abstract}
\abstractext
\end{abstract}
\end{@twocolumnfalse}
]

\section{Introduction}
\label{sec:intro}

\textit{Ab initio} electronic structure methods\cite{Szabo:1982,Helgaker:2000,Fetter2003,Dickhoff2008} are often used in combination with experiment to gain a deeper understanding of the excited-state properties of molecules and to study the behavior of these molecules as they undergo various photochemical transformations. Among many quantum chemical methods, algebraic diagrammatic construction theory\cite{Schirmer:1982p2395,Schirmer:1991p4647,Mertins:1996p2140,Schirmer:2004p11449,Dreuw:2014p82} (ADC) has become one of the widely used approaches for studying electronically excited states,\cite{Starcke:2009p024104,Harbach:2014p064113,Schirmer:1998p4734,Trofimov:2005p144115,Angonoa:1987p6789,Schirmer:2001p10621,Thiel:2003p2088,Barth:1985p867,Wenzel:2014p1900,Knippenberg:2012p064107,Pernpointner:2014p084108,Krauter:2017p286,Pernpointner:2018p1510,Dempwolff:2019p064108,Banerjee:2019p224112,Banerjee:2021p074105} due to a combination of its low computational cost and systematically improvable accuracy. In its original formulation proposed almost 40 years ago,\cite{Schirmer:1982p2395} ADC is based on the conventional (single-reference) perturbation theory, which significantly simplifies its working equations and allows for efficient calculation of excitation energies and excited-state properties. However, due to its single-reference nature, conventional ADC approximations become unreliable in systems with strong electron correlation in the ground or excited electronic states. 

Recently, we proposed a multireference formulation of ADC\cite{Sokolov:2018p204113} (MR-ADC) that can be considered as a natural generalization of the conventional ADC theory for strongly-correlated (multiconfigurational) wavefunctions. MR-ADC combines the advantages of the multireference perturbation theories,\cite{Wolinski:1987p225,Hirao:1992p374,Werner:1996p645,Finley:1998p299,Andersson:1990p5483,Andersson:1992p1218,Angeli:2001p10252,Angeli:2001p297,Angeli:2004p4043,Kurashige:2011p094104,Kurashige:2014p174111,Guo:2016p1583,Sokolov:2016p064102,Sharma:2017p488,Yanai:2017p4829,Sokolov:2017p244102} effective Hamiltonian approaches,\cite{Chattopadhyay:2000p7939,Chattopadhyay:2007p1787,Jagau:2012p044116,Samanta:2014p134108,Kohn:2019p041106,Datta:2012p204107,Nooijen:2014p081102,Huntington:2015p194111,Lechner:2021pe1939185} and multiconfigurational propagator methods\cite{Banerjee:1978p389,Yeager:1979p77,Dalgaard:1980p816,Yeager:1984p85,Graham:1991p2884,Yeager:1992p133,Nichols:1998p293,Khrustov:2002p507,HelmichParis:2019p174121,HelmichParis:2021pe26559} in a single theoretical framework that offers a hierarchy of computationally efficient approximations that provide a straightforward access to size-consistent excitation energies and excited-state properties and allow for calculation of excitations outside of strongly-correlated (active) orbitals. More recent work in our group concentrated on the development of the second-order MR-ADC approximations for simulating charged excitations (such as electron attachment or ionization) and demonstrated that the MR-ADC methods perform very well in a variety of weakly- and strongly-correlated systems.\cite{Chatterjee:2019p5908,Chatterjee:2020p6343} On the other hand, our computer implementation of MR-ADC for neutral electronic excitations has been limited to the first-order MR-ADC approximation (MR-ADC(1)),\cite{Sokolov:2018p204113} which did not provide accurate excitation energies missing much of the dynamic correlation effects.

In this work, we present the implementation and benchmark of the strict and extended second-order MR-ADC methods (MR-ADC(2) and MR-ADC(2)-X) for neutral electronic excitations that provide a more balanced description of static and dynamic electron correlation effects. We present an overview of the MR-ADC theory for neutral excitations in \cref{sec:theory}, before outlining the principle components of our implementation in \cref{sec:implementation}. We lay out the details of our computations in \cref{sec:computational_details} and present our benchmark results for excited states of small molecules (\ce{HF}, \ce{CO}, \ce{N2}, \ce{F2}, \ce{H2O}), the carbon dimer (\ce{C2}), as well as the ethylene (\ce{C2H4}) and butdiene (\ce{C4H6}) molecules in \cref{sec:results}. Our results demonstrate that the second-order MR-ADC approximations provide accurate predictions of the excited-state energies for weakly- and strongly-correlated electronic states at equilibrium geometries and near dissociation.  We outline our conclusions in \cref{sec:conclusions}. 

\section{Theory}
\label{sec:theory}
\subsection{MR-ADC for the Polarization Propagator}
\label{sec:theory:mr_adc_polarization_propagator}

We begin with a brief overview of the principal components of the MR-ADC theory in the context of neutral (particle-number-conserving) electronic excitations. A more detailed discussion of MR-ADC using the formalism of the effective Liouvillean theory\cite{Mukherjee:1989p257} can be found in Ref.\@ \citenum{Sokolov:2018p204113}. Electronic excitations arise from the perturbation of a molecule by an external electric field. The response of the molecule is described by the spectral function\cite{Schirmer:1982p2395}
\begin{equation}
	\label{eq:spectralfunction}
	T(\om) = -\frac{1}{\pi} \operatorname{Tr} \left[ \operatorname{Im} (\mathbf{D^{\dag}} \mathbf{\Pi}(\om) \mathbf{D}) \right]
\end{equation}
where $\mathbf{D}$ is a dipole moment operator matrix with elements $D_{pq}$ expressed in a finite single-particle basis set and $\mathbf{\Pi}(\om)$ is a polarization propagator, which is the central object of interest in this work.

The polarization propagator $\mathbf{\Pi}(\om)$ is a particular type of the retarded frequency-dependent propagator\cite{Fetter2003,Dickhoff2008} $\mathbf{G}(\om)$, which can be generally written as:
\begin{align}
	\label{eq:g_munu}
	G_{\mu\nu}(\omega)
	& = G_{\mu\nu}^+(\omega) \pm G_{\mu\nu}^-(\omega) \notag \\
	& =  \bra{\Psi}q_\mu(\omega - H + E)^{-1}q^\dag_\nu\ket{\Psi} \notag \\
	&\pm \bra{\Psi}q^\dag_\nu(\omega + H - E)^{-1}q_\mu\ket{\Psi}
\end{align}
where $G_{\mu\nu}^+(\omega)$ and $G_{\mu\nu}^-(\omega)$ are the forward and backward components of the propagator, $H$ is the electronic Hamiltonian, $\ket{\Psi}$ and $E$ are the eigenfunction and eigenenergy of $H$, respectively. The polarization propagator $\mathbf{\Pi}(\om)$ is characterized by the choice of $q^\dag_\nu = \c{p}\a{q} - \bra{\Psi}\c{p}\a{q}\ket{\Psi}$, expressed in terms of the fermionic annihilation and creation operators, and the negative sign for the second term in \cref{eq:g_munu}. In the case of the polarization propagator, the forward and backward components are related as $\mathbf{\Pi}_+^\dag(-\omega) = -\mathbf{\Pi}_-(\omega)$, so in computing one component, the other is trivially obtained. For this reason, in this work we define $\mathbf{\Pi}(\omega) \equiv \mathbf{\Pi}_+(\omega)$ and consider only the forward component of the propagator henceforth.

When written in the equivalent Lehmann\cite{Lehmann:1954bm} (spectral) representation, in which a resolution of identity is introduced over all excited eigenstates of the system ($\ket{\Psi_k}$),
\begin{align}
	\label{eq:lehmann}
	\Pi_{pqrs}(\om) = \sum_{k\neq0}   \frac{\braket{\Psi | \c{q}\a{p} | \Psi_k} \braket{\Psi_k | \c{r} \a{s} | \Psi}}   {\om + E_0 - E_k}.
\end{align}
the poles of the polarization propagator $\Pi_{pqrs}(\om)$ correspond to excitation energies ($\om_k = E_k  - E_0$) while the expectation values in the numerator describe the probabilities of the corresponding transitions. The frequency $\om \equiv \om' + i\eta$ is defined in terms of real ($\om'$) and infinitesimal imaginary ($i\eta$) components. \cref{eq:lehmann} can be written compactly in a matrix form
\begin{align}
	\label{eq:lehmannmatrix}
	\mathbf{\Pi}(\om) = \mathbf{\tilde{X}} (\om - \mathbf{\tilde{\Omega}})^{-1}  \mathbf{\tilde{X}}^\dag
\end{align}
where $\mathbf{\tilde{\Omega}}$ is a diagonal matrix of excitation energies ($\om_k$) and $\mathbf{\tilde{X}}$ is a matrix of spectroscopic amplitudes $\tilde{X}_{pqk} = \braket{\Psi | \c{q}\a{p} | \Psi_k}$. The expressions for the polarization propagator in \cref{eq:lehmann,eq:lehmannmatrix} are exact if written in the basis of the exact eigenstates, however in practice these states must be approximated.

\begin{figure}[t!]
	\includegraphics[width=0.5\textwidth]{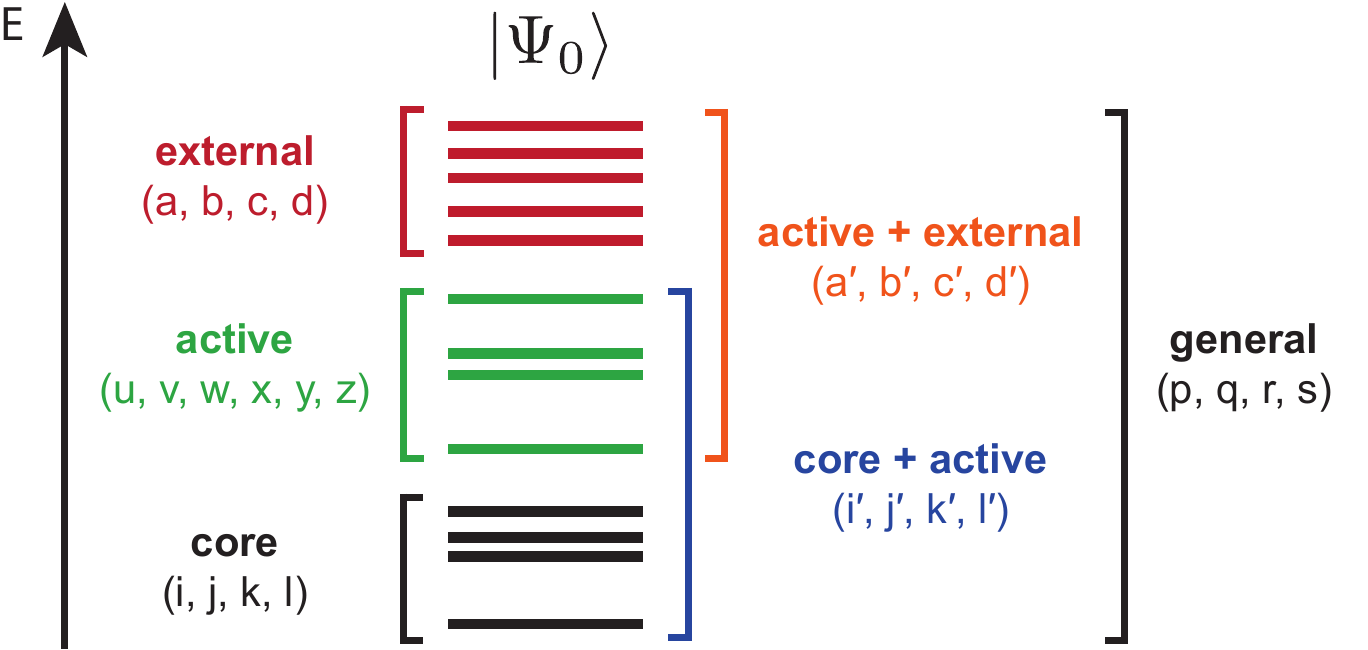}
	\captionsetup{justification=raggedright,singlelinecheck=false}
	\caption{Orbital indexing convention used in this work.}
	\label{fig:mo_diagram}
\end{figure}

MR-ADC provides a hierarchy of efficient approximations to the polarization propagator \eqref{eq:lehmannmatrix} by expressing it in a time-independent multireference perturbative expansion and truncating this series at a low order in perturbation theory. Working equations of the MR-ADC approximations are derived using the multireference generalization of the effective Liouvillean theory\cite{Sokolov:2018p204113,Mukherjee:1989p257}, which starts by writing the wavefunction $\ket{\Psi}$ as a unitary transformation\cite{Kirtman:1981p798,Hoffmann:1988p993,Mukherjee:1989p257,Yanai:2006p194106,Chen:2012p014108,Li:2015p2097} of the complete active-space self-consistent field\cite{Malmqvist:1989p189} (CASSCF) reference wavefunction $\ket{\Psi_0}$ 
\begin{align}
	\label{eq:mr_adc_wfn}
	\ket{\Psi} &= e^{A}\ket{\Psi_0}=e^{T-T^\dag}\ket{\Psi_0} , \quad T = \sum_{k=1}^N T_k \\
	\label{eq:mr_adc_t_amplitudes}
	T_k &= \frac{1}{(k!)^2} {\sum_{i'j'a'b'\ldots}} t_{i'j'\ldots}^{a'b'\ldots} \c{a'}\c{b'}\ldots\a{j'}\a{i'}
\end{align}
In \cref{eq:mr_adc_wfn,eq:mr_adc_t_amplitudes}, $T$ is an excitation operator that generates all of the internally-contracted excitations between the core, active, and external orbitals (\cref{fig:mo_diagram}). Next, the full electronic Hamiltonian $H$ is partitioned into the zeroth-order part $H^{(0)}$ and a perturbation $V = H - H^{(0)}$, where $H^{(0)}$ is the Dyall Hamiltonian\cite{Dyall:1995p4909,Angeli:2001p10252,Angeli:2001p297,Angeli:2004p4043} with an eigenfunction $\ket{\Psi_0}$. This partitioning of $H$ leads to the perturbative expansions for the wavefunction $\ket{\Psi}$ and the polarization propagator $\mathbf{\Pi}(\om)$:
\begin{align}
	\ket{\Psi} &= e^{A^{(0)} + A^{(1)}+\cdots+A^{(n)}+\cdots}\ket{\Psi_0} \\
	\mathbf{\Pi}(\om) &= \mathbf{\Pi}^{(0)}(\om) + \mathbf{\Pi}^{(1)}(\om) +  \cdots  + \mathbf{\Pi}^{(n)}(\om) + \cdots
\end{align}
Truncating these expansions at the \textit{n}\textsuperscript{th} order defines the polarization propagator of the MR-ADC(\textit{n}) approximation. The MR-ADC(\textit{n}) working equations are derived by writing $\mathbf{\Pi}(\om)$ in the form
\begin{align}
	\label{eq:ADCprop}
	\mathbf{\Pi}(\omega) = \mathbf{T} (\omega \mathbf{S} - \mathbf{M})^{-1} \mathbf{T}^{\dag}
\end{align}
where $\mathbf{M}$ is a non-diagonal effective Hamiltonian matrix that can be used to compute the excitation energies and $\mathbf{T}$ is an effective transition moments matrix that contains information about the probabilities of electronic transitions. The matrices $\mathbf{M}$ and $\mathbf{T}$ are evaluated in an approximate basis of many-electron (internally-contracted) states, which are in general non-orthogonal with an overlap matrix $\mathbf{S}$. In \cref{eq:ADCprop} each matrix is evaluated up to the same (\textit{n}\textsuperscript{th}) order in perturbation theory.

The MR-ADC(n) excitation energies $\mathbf{\Omega}$ and complementary eigenvectors $\mathbf{Y}$ are obtained by solving the generalized eigenvalue problem
\begin{align}
	\label{eq:eigenprob}
	\mathbf{M} \mathbf{Y} = \mathbf{S} \mathbf{Y} \mathbf{\Omega}
\end{align}
Combining the eigenvectors $\mathbf{Y}$ with the effective transition moments matrix $\mathbf{T}$ allows to compute the spectroscopic amplitudes
\begin{align}
	\label{eq:spec_amp}
	\mathbf{X} = \mathbf{T} \mathbf{S}^{-\frac{1}{2}} \mathbf{Y}
\end{align}
and the MR-ADC(n) polarization propagator
\begin{align}
	\label{eq:ADC_diag}
	\mathbf{\Pi}(\om) = \mathbf{X} (\om - \mathbf{\Omega})^{-1}  \mathbf{X}^\dag
\end{align}
along with the spectral function $T(\om)$ in \cref{eq:spectralfunction}. The spectroscopic amplitudes $\mathbf{X}$ can be also used to compute the oscillator strength for each electronic transition $k$
\begin{align}
	\label{eq:osc_strength}
	f_k = \frac{2}{3} \omega_k \left|\braket{\Psi | D | \Psi_k}\right|^2 = \frac{2}{3} \omega_k \left(\sum_{pq} D_{pq} X_{pqk}\right)^2
\end{align}
where $\braket{\Psi | D | \Psi_k}$ is the transition dipole moment matrix element. 

\subsection{Strict and Extended Second-Order MR-ADC Approximations}
\label{sec:theory:mr_adc2_x}
\subsubsection{Perturbative Structure of the MR-ADC Matrices}
\label{sec:theory:mr_adc2_x:pt}

In this work, we consider the strict and extended second-order MR-ADC approximations for electronic excitations, denoted as MR-ADC(2) and MR-ADC(2)-X. In the MR-ADC(2) approximation, contributions to $\mathbf{\Pi}(\om)$ are included strictly up to the second order in the MR-ADC perturbation expansion, while the MR-ADC(2)-X method partially incorporates selected third-order terms as discussed below. 

\begin{figure*}[t!]
	\includegraphics[width=0.9\textwidth]{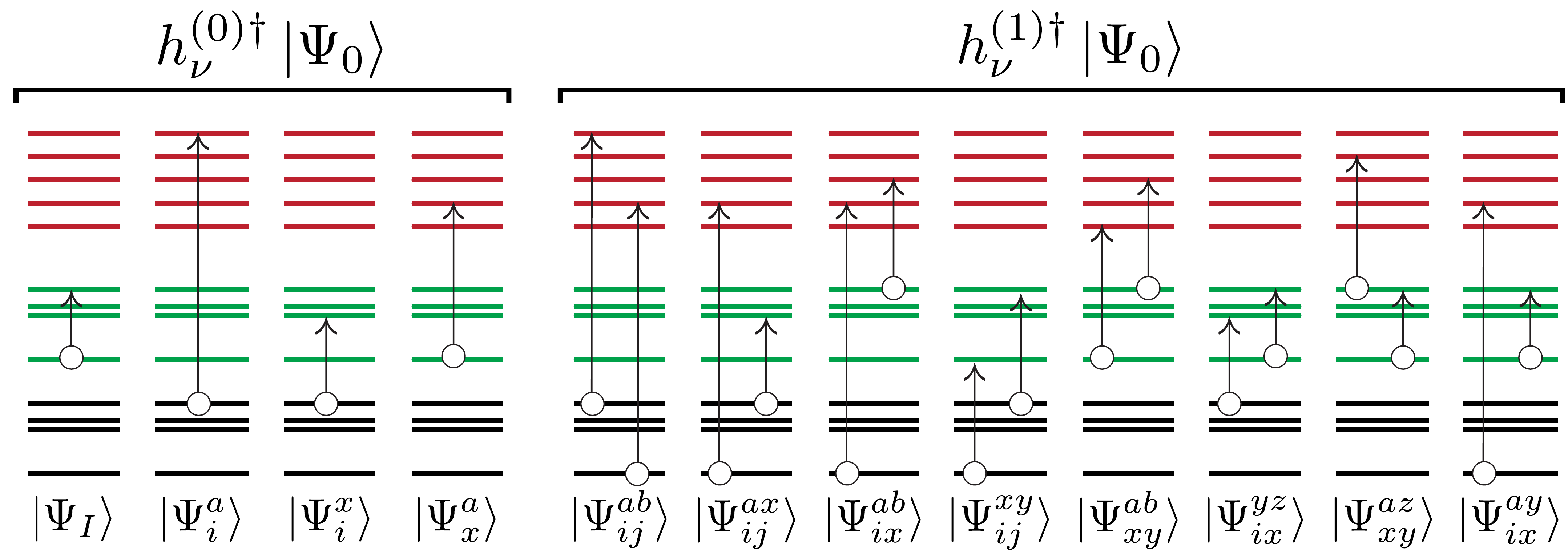}
	\captionsetup{justification=raggedright,singlelinecheck=false}
	\caption{Schematic illustration of the electronically-excited basis states produced by acting the $h_{\nu}^{(k)\dag}$ ($k = 0, 1$) operators (\cref{eq:h_manifold_0,eq:h_manifold_1}) on the reference state $\ket{\Psi_0}$. A circle connected with an arrow denotes a single excitation. Black, green, and red lines represent core, active, and external orbitals, respectively.}
	\label{fig:h_manifold}
\end{figure*}

At each perturbation order \textit{n}, contributions to the matrix elements of $\mathbf{M}$, $\mathbf{T}$, and $\mathbf{S}$ in \cref{eq:ADCprop} are expressed as
\begin{align}
	\label{eq:M_elements}
	M^{(n)}_{\mu\nu} &= \sum^{k + l + m = n}_{klm}\braket{\Psi_0 | [h^{(k)}_\mu, [\tilde{H}^{(l)}, h^{(m)\dag}_\nu]] | \Psi_0} \\
	\label{eq:T_elements}
	T^{(n)}_{\mu\nu} &= \sum^{k + m = n}_{km}\braket{\Psi_0 | [\tilde{q}^{(k)}_\mu, h^{(m)\dag}_\nu] | \Psi_0} \\
	\label{eq:S_elements}
	S^{(n)}_{\mu\nu} &= \sum^{k + m = n}_{km}\braket{\Psi_0 | [h^{(k)}_\mu, h^{(m)\dag}_\nu] | \Psi_0}
\end{align} 
where $\tilde{H}^{(l)}$ and $\tilde{q}^{(l)}_\mu$ are the $\mathit{l}$\textsuperscript{th}-order operators obtained from the perturbative analysis of the Baker--Campbell--Hausdorff expansions for the effective Hamiltonian $\tilde{H} = e^{-A}He^{A}$ and observable $\tilde{q}_\mu = e^{-A} q_\mu e^{A}$ operators, respectively. The excitation operators $h^{(m)\dag}_\nu$ are used to construct the many-electron basis states that are necessary for representing the wavefunctions of the excited electronic states. Defining $a^p_q \equiv \c{p} \a{q}$ and $a^{pq}_{rs} \equiv \c{p} \c{q} \a{s} \a{r}$, the low-order $h^{(m)\dag}_\nu$ used in MR-ADC(2) and MR-ADC(2)-X have the form:
\begin{align}
	\mathbf{h}^{(0)\dag} &= \{Z^\dag_I \ (I > 0),\ a^a_i,\  a^x_i,\  a^a_x\} \label{eq:h_manifold_0} \\
	\mathbf{h}^{(1)\dag} &= \{a^{ab}_{ij},\  a^{ax}_{ij},\  a^{ab}_{ix},\  a^{xy}_{ij},\  a^{ab}_{xy},\  a^{yz}_{ix},\  a^{az}_{xy},\  a^{ay}_{ix}\} \label{eq:h_manifold_1}
\end{align}
These excitations are pictorially represented in \cref{fig:h_manifold}. The $\mathbf{h}^{(0)\dag}$ operator set contains two classes of operators: (i) the eigenoperators\cite{Lowdin:1985p285,Kutzelnigg:1998p5578} $Z^\dag_I = \ket{\Psi_I}\bra{\Psi_0}$ $(I>0)$ that describe excitations within the active space and (ii) the operators $a^{p}_{q}$ that produce single excitations outside of the active space. The excited-state active-space wavefunctions $\ket{\Psi_I}$ $(I>0)$ that appear in $Z^\dag_I$ are obtained by performing a multistate CASCI calculation using the CASSCF orbitals optimized for the reference state $\ket{\Psi_0}$. Although, formally, the set of $\ket{\Psi_I}$ $(I>0)$ includes all CASCI wavefunctions in a given complete active space, in practice only a small number of CASCI states relevant to the spectral region of interest need to be computed. As shown in \cref{fig:h_manifold}, the $\mathbf{h}^{(1)\dag}$ operator set describes all possible double excitations outside of the active space.

\begin{figure*}[t!]
	\subfloat[]{\label{fig:M_T_strict}\includegraphics[width=0.45\textwidth]{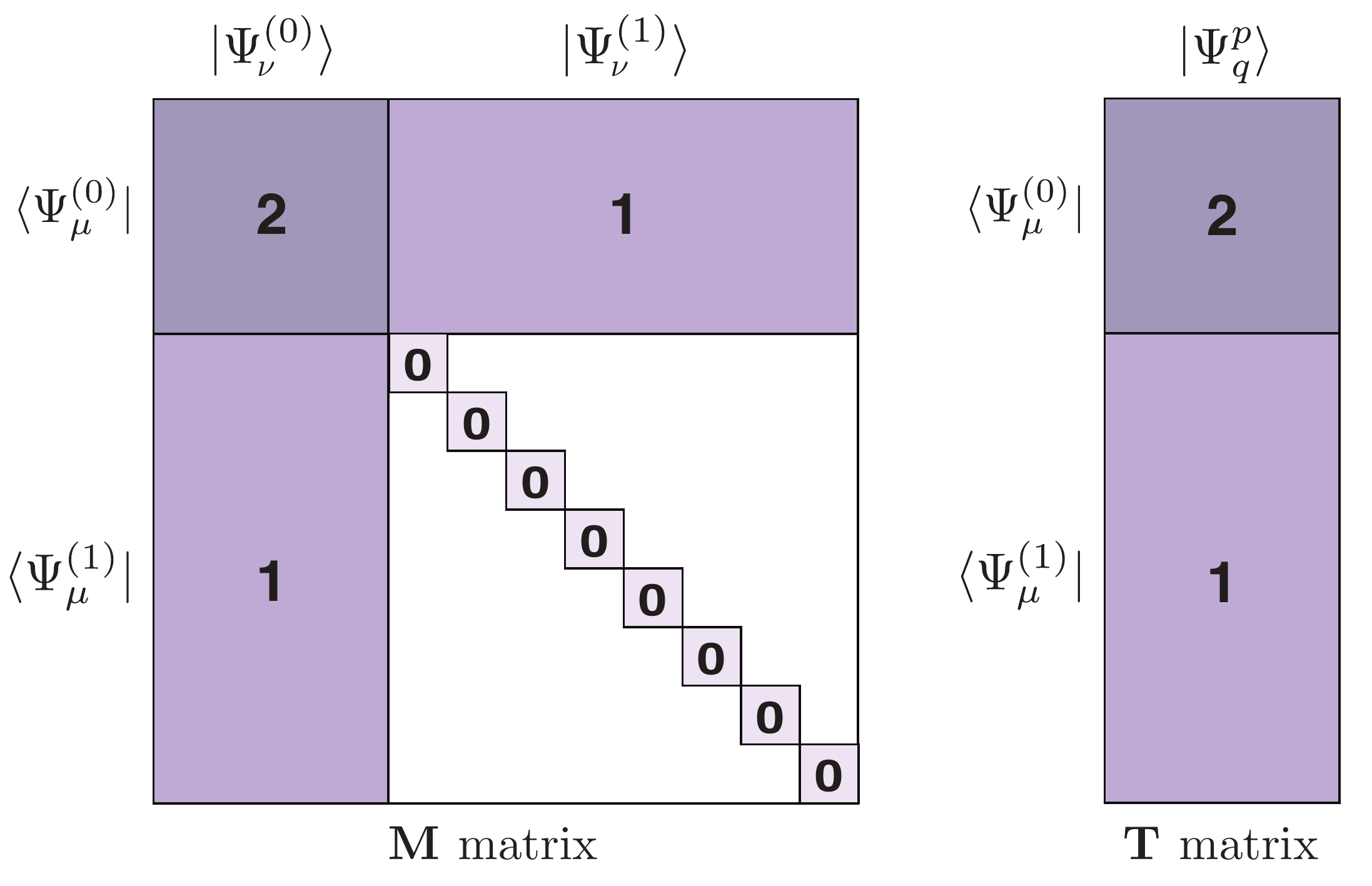}} \quad
	\subfloat[]{\label{fig:M_T_extended}\includegraphics[width=0.45\textwidth]{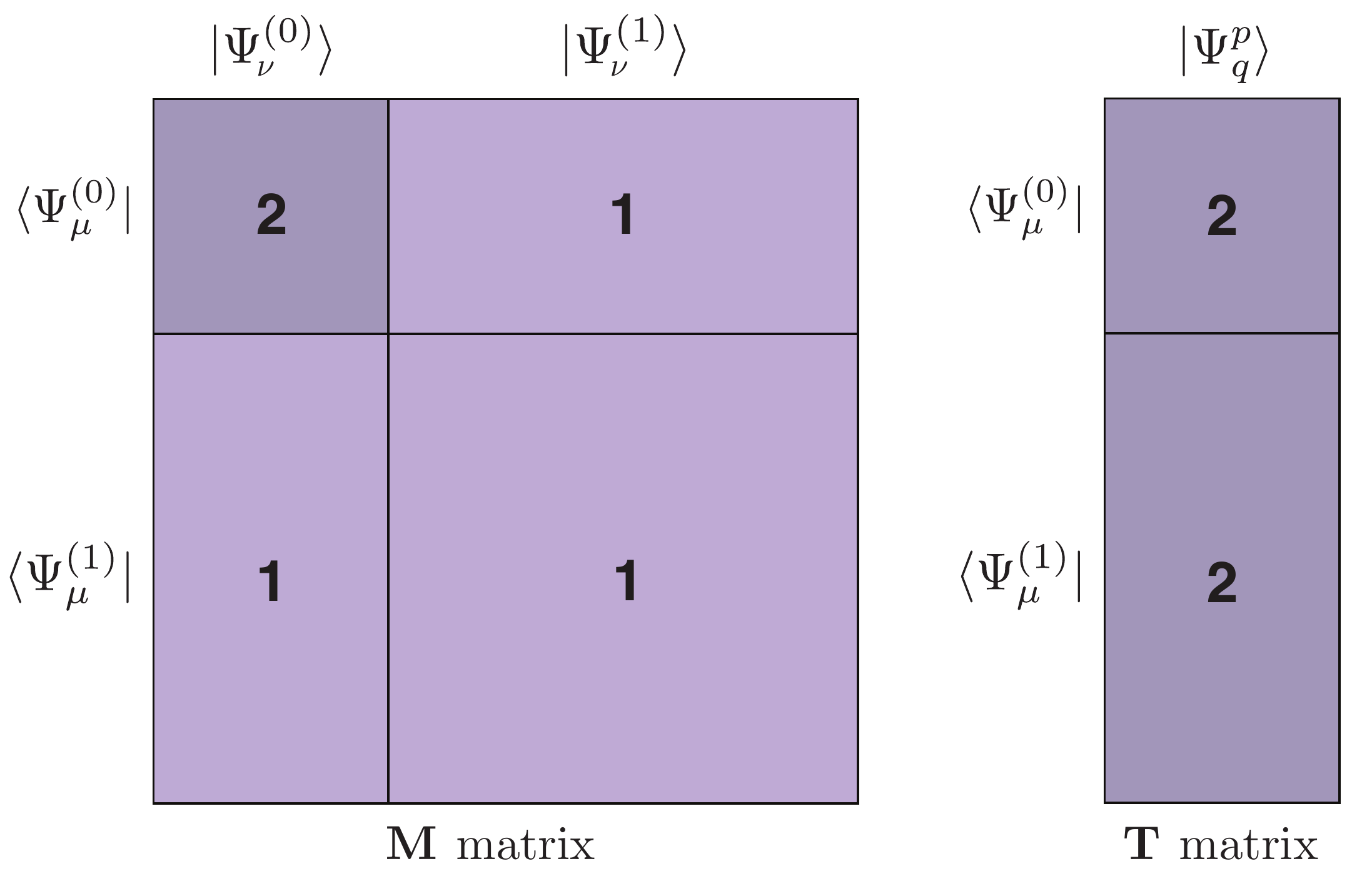}}
	\captionsetup{justification=raggedright,singlelinecheck=false}
	\caption{Graphical representation of the effective Hamiltonian $\mathbf{M}$ and transition moments $\mathbf{T}$ matrices for the a) MR-ADC(2) and b) MR-ADC(2)-X methods.}
	\label{fig:M_T_matrices}
\end{figure*}

\cref{eq:M_elements,eq:T_elements} define the perturbative structure of the $\mathbf{M}$ and $\mathbf{T}$ matrices shown in \cref{fig:M_T_matrices}, with the effective operators  $\tilde{H}^{(k)}$ and $\tilde{q}^{(l)}$ expanded to different orders in different sectors defined by the $h^{(0)\dag}_\nu$ and $h^{(1)\dag}_\nu$ operators. Both MR-ADC(2) and MR-ADC(2)-X incorporate up to $\tilde{H}^{(2)}$ in the $\braket{\Psi_0 | [h^{(0)}_\mu, [\tilde{H}^{(l)}, h^{(0)\dag}_\nu]] | \Psi_0}$ sector of the $\mathbf{M}$ matrix and up to $\tilde{H}^{(1)}$ in the $h^{(0)}_\mu$ -- $h^{(1)\dag}_\nu$ and $h^{(1)}_\mu$ -- $h^{(0)\dag}_\nu$ coupling blocks. The two methods differ in the $h^{(1)}_\mu$ -- $h^{(1)\dag}_\nu$ sector approximating the effective Hamiltonian to $\tilde{H}^{(0)}$ and $\tilde{H}^{(1)}$ in MR-ADC(2) and MR-ADC(2)-X, respectively. For the $\mathbf{T}$ matrix, MR-ADC(2) and MR-ADC(2)-X describe the effective observable operator up to $\tilde{q}^{(2)}$ in the $h^{(0)\dag}_\nu$ sector, but differ in its expansion in the $h^{(1)\dag}_\nu$ block (\cref{fig:M_T_matrices}). We note that the perturbative structure of the $\mathbf{M}$ and $\mathbf{T}$ matrices in MR-ADC is very similar to that in single-reference ADC\cite{Dreuw:2014p82} and the two theories become equivalent when the number of active orbitals is zero.

\subsubsection{Amplitudes of the Effective Hamiltonian}
\label{sec:theory:T_amps}

Computation of the $\mathbf{M}$ and $\mathbf{T}$ matrix elements in \cref{fig:M_T_matrices} requires solving for the amplitudes of the excitation operator  $T^{(k)}$ (\cref{eq:mr_adc_t_amplitudes}). We express $T^{(k)}$ in a general form as
\begin{align}
	\label{eq:excit_op_tensor}
	T^{(k)} = \mathbf{t^{(k)}} \, \boldsymbol{\tau}^\dag =  \sum_\mu t_\mu^{(k)} \tau^\dag_\mu
\end{align}
where $t_\mu^{(k)}$ are the kth-order amplitude coefficients and $\tau^\dag_\mu$ denotes the corresponding string of creation and annihilation operators. In MR-ADC(2) and MR-ADC(2)-X only the low-order amplitudes $\mathbf{t^{(1)}}$ and $\mathbf{t^{(2)}}$ need to be evaluated. These amplitudes are determined by solving a system of projected linear equations
\begin{align}
	\label{eq:proj_amplitude_equations}
	\braket{\Psi_0|\tau_\mu\tilde{H}^{(k)}|\Psi_0} = 0 \qquad (k = 1,\ 2)
\end{align}
For the details of solving these amplitude equations we refer the readers to our earlier publications.\cite{Chatterjee:2019p5908,Chatterjee:2020p6343}

Both MR-ADC(2) and MR-ADC(2)-X require calculating all first-order amplitudes, specifically:
\begin{align}
	\label{eq:t1_amp_tensor}
	\mathbf{t^{(1)}} = &\left\{t_{i}^{a(1)};\ t_{i}^{x(1)};\ t_{x}^{a(1)};\ t_{ij}^{ab(1)};\ t_{ij}^{ax(1)};\ t_{ix}^{ab(1)}; \right. \notag \\
	&\left. t_{ij}^{xy(1)};\ t_{xy}^{ab(1)};\ t_{ix}^{ay(1)};\ t_{ix}^{yz(1)};\ t_{xy}^{az(1)}\right\}
\end{align}
The first three sets of amplitudes correspond to single excitations ($t_{i}^{a(1)};\ t_{i}^{x(1)};\ t_{x}^{a(1)}$) and the next five sets represent external double excitations. The last three sets of $\mathbf{t^{(1)}}$ in \cref{eq:t1_amp_tensor} describe the semi-internal double excitations ($t_{ix}^{ay(1)};\ t_{ix}^{yz(1)};\ t_{xy}^{az(1)}$).

The MR-ADC(2) and MR-ADC(2)-X equations for the effective Hamiltonian matrix  $\mathbf{M}$ also depend on the second-order singles and semi-internal doubles amplitudes of the form:
\begin{align}
	\label{eq:t2_amp_tensor}
	\mathbf{t^{(2)}_{s}} = &\left\{t_{i}^{a(2)};\ t_{i}^{x(2)};\ t_{x}^{a(2)};\ t_{x}^{y(2)} (x>y);\right. \notag\\
	&\left. t_{ix}^{ay(2)};\ t_{ix}^{yz(2)};\ t_{xy}^{az(2)}\right\}
\end{align}
We note that in addition to the external singles amplitudes ($t_{i}^{a(2)};\ t_{i}^{x(2)};\ t_{x}^{a(2)}$), \cref{eq:t2_amp_tensor} also includes the internal singles amplitudes $t_{x}^{y(2)} (x>y)$ that we did not encounter in our previous work. These amplitudes are necessary to ensure the Hermiticity of the effective Hamiltonian matrix $\mathbf{M}$ as discussed below. Additionally, the MR-ADC(2)-X equations for the $\mathbf{T}$ matrix elements (\cref{fig:M_T_extended}) depend on the external double-excitation amplitudes
\begin{align}
	\label{eq:t2_amp_tensor_X}
	\mathbf{t^{(2)}_{d}} = &\left\{t_{ij}^{ab(2)};\ t_{ij}^{ax(2)}; t_{ix}^{ab(2)};\ t_{ij}^{xy(2)};\ t_{xy}^{ab(2)}\right\}
\end{align}

In our previous work on EA/IP-MR-ADC(2) and EA/IP-MR-ADC(2)-X,\cite{Chatterjee:2019p5908,Chatterjee:2020p6343} we demonstrated that neglecting all second-order amplitudes except $t_{i}^{a(2)}$ has a negligible effect on the computed charged excitation energies and transition probabilities. However, for neutral electronic excitations, some of the second-order amplitudes need to be computed to prevent asymmetry of the $\mathbf{M}$ matrix. To illustrate this, we consider the second-order contributions to one of the off-diagonal blocks of $\mathbf{M}$ and its transpose that can be expressed and simplified as: 
\begin{align}
	\label{eq:M_asym}
	M^{(2)}_{ix,jb} = &\braket{\Psi_0 | [\c{i} \a{x}, [\tilde{H}^{(2)}, \c{b} \a{j}]] | \Psi_0}   \notag \\
				= &\braket{\Psi_0 | \c{i} \a{x} \tilde{H}^{(2)} \c{b} \a{j} | \Psi_0}  \\
	\label{eq:M_asym_transpose}
	M^{(2)\dag}_{ix,jb} = &\braket{\Psi_0 | [\c{j} \a{b}, [\tilde{H}^{(2)}, \c{x} \a{i}]] | \Psi_0}^\dag   \notag \\
				= &\braket{\Psi_0 | \c{i} \a{x} \tilde{H}^{(2)}  \c{b} \a{j}  | \Psi_0}  \notag \\
                + &\d{i}{j} \braket{\Psi_0 |  \tilde{H}^{(2)} \c{b} \a{x}  | \Psi_0}
\end{align}
The two blocks of the $\mathbf{M}$ matrix are equal to each other provided that the second term in \cref{eq:M_asym_transpose} is zero, which is satisfied if the second-order effective Hamiltonian $\tilde{H}^{(2)}$ is parametrized with the $t_{x}^{b(2)}$ amplitudes that fulfill the projected amplitude equations $\braket{\Psi_0 |  \tilde{H}^{(2)} \c{b} \a{x}  | \Psi_0} = 0$ (\cref{eq:proj_amplitude_equations}). As another example we consider one of the diagonal blocks of $\mathbf{M}$:
\begin{align}
	\label{eq:M_asym_xy}
	M^{(2)}_{ix,jy} = &\braket{\Psi_0 | [\c{i} \a{x}, [\tilde{H}^{(2)}, \c{y} \a{j}]] | \Psi_0}   \notag \\
				= &\braket{\Psi_0 | \c{i} \a{x} \tilde{H}^{(2)} \c{y} \a{j} | \Psi_0}  \notag \\
                - &\d{i}{j} \braket{\Psi_0 |  \a{x} \c{y} \tilde{H}^{(2)} | \Psi_0} \\
	\label{eq:M_asym_xy_transpose}
	M^{(2)\dag}_{ix,jy} = &\braket{\Psi_0 | [\c{j} \a{y}, [\tilde{H}^{(2)}, \c{x} \a{i}]] | \Psi_0}^\dag   \notag \\
				= &\braket{\Psi_0 | \c{i} \a{x} \tilde{H}^{(2)}  \c{y} \a{j}  | \Psi_0}  \notag \\
                - &\d{i}{j} \braket{\Psi_0 | \tilde{H}^{(2)}  \a{x}  \c{y} | \Psi_0}
\end{align}
Here, the last two terms in \cref{eq:M_asym_xy,eq:M_asym_xy_transpose} are in general different, unless the effective Hamiltonian includes contributions from the internal second-order amplitudes $t_{x}^{y(2)} (x>y)$ ensuring that $\braket{\Psi_0 |  \a{x} \c{y} \tilde{H}^{(2)} | \Psi_0} = \braket{\Psi_0 | \tilde{H}^{(2)}  \a{x}  \c{y} | \Psi_0}$, which results in the Hermitian $M^{(2)}_{ix,jy}$ block. The solution of the amplitude equations for the $t_{x}^{y(2)} (x>y)$ amplitudes is described in the Appendix.

Including all second-order singles amplitudes ensures the full Hermiticity of the $\mathbf{M}$ matrix in all sectors. On the other hand, neglecting the second-order doubles amplitudes does not affect the symmetry of $\mathbf{M}$ and has a negligible effect on the computed excitation energies and oscillator strengths (see Table S1 of the Supporting Information for details). For this reason, in our implementation of MR-ADC for electronic excitations, we approximate:
\begin{align}
	\label{eq:t2_amp_tensor_approx}
	\mathbf{t^{(2)}} \approx &\left\{t_{i}^{a(2)};\ t_{i}^{x(2)};\ t_{x}^{a(2)};\ t_{x}^{y(2)} (x>y) \right\}
\end{align}

\section{Implementation}
\label{sec:implementation}

Our implementation of MR-ADC(2) and MR-ADC(2)-X for electronic excitations follows the steps outlined below:

\begin{enumerate}
	\item Given an active space, compute the reference (usually, ground-state) CASSCF wavefunction $\ket{\Psi_0}$ and molecular orbitals. 
	\item Using the reference CASSCF molecular orbitals compute the CASCI wavefunctions $\ket{\Psi_I}\, (I > 0)$ for the user-defined number of excited states ($N_{\mathrm{CI}}$).
	\item Calculate the reduced density matrices (RDMs):
	\begin{enumerate}
		\item \textit{reference} RDMs with respect to $\ket{\Psi_0}$;
		\item \textit{transition} RDMs between $\ket{\Psi_0}$ and $\ket{\Psi_I}\, (I > 0)$;
		\item \textit{excited-state} RDMs amongst $\ket{\Psi_I}\, (I > 0)$.
	\end{enumerate}
	\item Solve the linear amplitude equations (\cref{eq:proj_amplitude_equations}) to compute the first- and second-order amplitudes $\mathbf{t^{(1)}}$ and $\mathbf{t^{(2)}}$ discussed in \cref{sec:theory:T_amps}. 
	\item Diagonalize the effective Hamiltonian matrix $\mathbf{M}$ (\cref{eq:eigenprob}) to compute excitation energies.
	\item Compute the spectroscopic amplitudes $\mathbf{X}$ (\cref{eq:spec_amp}) and oscillator strengths $f_k$ (\cref{eq:osc_strength}).
\end{enumerate}	

As discussed in our previous work,\cite{Chatterjee:2019p5908,Chatterjee:2020p6343} the MR-ADC generalized eigenvalue problem is solved in the symmetrically-orthogonalized form ($\mathbf{\tilde{M}} \, \mathbf{\tilde{Y}} = \mathbf{\tilde{Y}} \, \mathbf{\Omega}$, $\mathbf{\tilde{M}} = \mathbf{S^{-1/2}} \, \mathbf{M} \, \mathbf{S^{-1/2}}$, $\mathbf{\tilde{Y}} = \mathbf{S^{1/2}} \, \mathbf{Y}$) using the multiroot Davidson algorithm\cite{Davidson:1975p87,Liu:1978p49} that iteratively optimizes the eigenvectors $\mathbf{\tilde{Y}}$ until convergence by forming the matrix-vector products $\boldsymbol{\sigma} = \mathbf{\tilde{M}\tilde{Y}}$. We note that the $\mathbf{S^{-1/2}}$ matrix has a block-diagonal structure that allows to avoid storing the full matrix in memory. The small non-diagonal blocks of $\mathbf{S^{-1/2}}$ are computed very efficiently by diagonalizing small sectors of the overlap matrix $\mathbf{S}$.\cite{Chatterjee:2019p5908} Equations and computer code for the calculation of the $\boldsymbol{\sigma}$ vectors were generated using a modified version of the \textsc{SecondQuantizationAlgebra} program (SQA) developed by Neuscamman and co-workers.\cite{Neuscamman:2009p124102} To optimize computational efficiency, our code generator implements the tensor contractions by automatically separating them into the efficient tensor intermediates, which are reused throughout the calculation.

For a fixed active space, the MR-ADC(2) and MR-ADC(2)-X implementations have the  $\mathcal{O}(M^5)$ and  $\mathcal{O}(M^6)$ computational scaling with the basis set size $M$, respectively, which is equivalent to the scaling of their single-reference counterparts.\cite{Dreuw:2014p82} Both methods have the formal $\mathcal{O}(N_{\mathrm{det}} N^8_{\mathrm{act}})$ scaling with the number of active orbitals ($N_{\mathrm{act}}$) and Slater determinants ($N_{\mathrm{det}}$), similar to that of the conventional second-order multireference perturbation theories. The active-space scaling can be further lowered to $\mathcal{O}(N_{\mathrm{det}} N^6_{\mathrm{act}})$ using a technique described in our previous work,\cite{Chatterjee:2019p5908,Chatterjee:2020p6343} which forms efficient intermediates to avoid computation of the four-particle RDMs, although we did not employ this approach in our current implementation. 

\section{Computational Details}
\label{sec:computational_details}

The MR-ADC(2) and MR-ADC(2)-X methods were implemented in \textsc{Prism}, a standalone program that is being developed in our group. The \textsc{Prism} program interfaces with \textsc{PySCF}\cite{Sun:2020p024109} to obtain integrals and CASSCF/CASCI reference wavefunctions. The MR-ADC results were benchmarked against excitation energies obtained from the semi-stochastic heat-bath configuration interaction (SHCI)\cite{Holmes:2016p3674,Sharma:2017p1595,Holmes:2017p164111} calculations extrapolated to the full configuration interaction limit. The SHCI method was implemented in the \textsc{Dice} program.\cite{Holmes:2016p3674,Sharma:2017p1595,Holmes:2017p164111} We also compare the MR-ADC results to those computed using strongly-contracted $N$-electron valence second-order perturbation theory (sc-NEVPT2),\cite{Angeli:2001p10252,Angeli:2001p297} single-reference ADC (ADC(n), n = 2, 3),\cite{Schirmer:1982p2395,Dreuw:2014p82} and equation-of-motion coupled cluster theory with single and double excitations (EOM-CCSD)\cite{Geertsen:1989p57,Comeau:1993p414,Stanton:1993p7029,Krylov:2008p433}. We used \textsc{PySCF} to obtain the sc-NEVPT2 energies. The ADC(2), ADC(3), and EOM-CCSD results were computed using \textsc{Q-Chem}\cite{qchem:44}.

Performance of the MR-ADC(2) and MR-ADC(2)-X methods was tested for a benchmark set of five small molecules (\ce{HF}, \ce{F2}, \ce{CO}, \ce{N2}, and \ce{H2O}), a carbon dimer (\ce{C2}), and two alkenes: ethylene and butadiene. As in our earlier work,\cite{Chatterjee:2019p5908,Chatterjee:2020p6343} computations of the five small molecules were performed using two geometries: equilibrium and stretched. The equilibrium geometries were taken from Ref.\@ \citenum{Trofimov:2005p144115}. The stretched geometries of all molecules were obtained by increasing their bond lengths by a factor of two. The C$-$C bond distance in the carbon dimer was set to 2.4 \bohr, which is close to its equilibrium bond length.\cite{Wouters:2014p1501} The geometries of the alkenes were obtained as described in Ref.\@ \citenum{Daday:2012p4441} and are reported in the Supporting Information. Calculations of \ce{HF}, \ce{F2}, \ce{CO}, \ce{N2}, and \ce{H2O} employed the aug-cc-pVDZ basis set.\cite{Kendall:1992p6796} Carbon dimer calculations were performed using cc-pVDZ.\cite{Dunning:1989p1007} For the ethylene and butadiene molecules we used the ANO-L-PVTZ and ANO-L-PVDZ basis sets, respectively.\cite{Widmark:1990p291}

\begin{table*}[t!]
	\caption{Vertical excitation energies ($\Omega$, eV) and oscillator strengths ($f$) of molecules with equilibrium geometries. See \cref{sec:computational_details} for details of the calculations. Also shown are mean absolute errors (\mae), standard deviations (\std), and maximum absolute errors (\maxe) of the energies, relative to SHCI. Only orbitals participating in excitations are listed in the electron configuration of each excited state.}
	\label{tab:ee_r_eq}
	\setlength{\extrarowheight}{2pt}
	\setstretch{1}
	\tiny
	\centering
	\hspace*{-0.8cm}
	\begin{threeparttable}
		\begin{tabular}{lrlccccccc}
			\hline\hline
			System   &                          Configuration & Term                 &        ADC(2)        &         ADC(3)         &      EOM-CCSD      & sc-NEVPT2 &        MR-ADC(2)         &        MR-ADC(2)-X        &   SHCI   \\
			&                                        &                      &     $\Omega \ (f)$     &      $\Omega \ (f)$      &      $\Omega$      & $\Omega$  &       $\Omega \ (f)$       &       $\Omega \ (f)$        & $\Omega$ \\
			\hline
			\ce{HF}  &       $(3\si)^{2}(1\pi)^{3}(4\si)^{1}$ & ${}^{3}\Pi$          &        $9.30$        &        $10.50$         &       $9.89$       &  $10.09$  &         $10.40$          &          $10.21$          & $10.05$  \\
			&       $(3\si)^{2}(1\pi)^{3}(4\si)^{1}$ & ${}^{1}\Pi$          & $9.63$    ($0.039$)  & $10.96$     ($0.042$)  &      $10.30$       &  $10.46$  & $10.80$       ($0.036$)  & $10.58$        ($0.035$)  & $10.44$  \\
			&       $(3\si)^{1}(1\pi)^{4}(4\si)^{1}$ & ${}^{3}\Sigma^{+}$   &       $13.05$        &        $13.80$         &      $13.41$       &  $13.60$  &         $13.91$          &          $13.67$          & $13.54$  \\
			&       $(3\si)^{2}(1\pi)^{3}(5\si)^{1}$ & ${}^{3}\Pi$          &       $13.14$        &        $14.53$         &      $13.86$       &  $14.06$  &         $14.17$          &          $13.96$          & $14.02$  \\
			&       $(3\si)^{2}(1\pi)^{3}(5\si)^{1}$ & ${}^{1}\Pi$          & $13.35$   ($0.006$)  & $14.74$     ($0.019$)  &      $14.06$       &  $14.27$  & $14.38$       ($0.010$)  & $14.17$        ($0.009$)  & $14.21$  \\
			&       $(3\si)^{2}(1\pi)^{3}(2\pi)^{1}$ & ${}^{3}\Sigma^{+}$   &       $13.80$        &        $14.83$         &      $14.26$       &  $14.52$  &         $14.70$          &          $14.52$          & $14.47$  \\
			&       $(3\si)^{1}(1\pi)^{4}(4\si)^{1}$ & ${}^{1}\Sigma^{+}$   & $13.91$   ($0.149$)  & $15.01$     ($0.178$)  &      $14.47$       &  $14.66$  & $15.08$       ($0.172$)  & $14.77$        ($0.167$)  & $14.58$  \\
			&       $(3\si)^{2}(1\pi)^{3}(2\pi)^{1}$ & ${}^{3}\Delta$       &       $14.22$        &        $15.33$         &      $14.74$       &  $14.93$  &         $15.18$          &          $14.99$          & $14.93$  \\
			&       $(3\si)^{2}(1\pi)^{3}(2\pi)^{1}$ & ${}^{1}\Delta$       &       $14.46$        &        $15.66$         &      $15.03$       &  $15.20$  &         $15.49$          &          $15.29$          & $15.20$  \\
			&       $(3\si)^{2}(1\pi)^{3}(2\pi)^{1}$ & ${}^{3}\Sigma^{-}$   &       $14.51$        &        $15.69$         &      $15.07$       &  $15.26$  &         $15.53$          &          $15.31$          & $15.25$  \\
			&       $(3\si)^{1}(1\pi)^{3}(2\pi)^{1}$ & ${}^{1}\Sigma^{-}$   &       $14.54$        &        $15.72$         &      $15.10$       &  $15.28$  &         $15.56$          &          $15.36$          & $15.28$  \\
			\hline
			\ce{CO}  &       $(1\pi)^{4}(5\si)^{1}(2\pi)^{1}$ & ${}^{3}\Pi$          &        $6.49$        &         $6.00$         &       $6.42$       &  $6.57$   &          $6.51$          &          $6.02$           &  $6.33$  \\
			&       $(1\pi)^{3}(5\si)^{2}(2\pi)^{1}$ & ${}^{3}\Sigma^{+}$   &        $8.60$        &         $8.38$         &       $8.44$       &  $8.67$   &          $8.94$          &          $8.81$           &  $8.54$  \\
			&       $(1\pi)^{4}(5\si)^{1}(2\pi)^{1}$ & ${}^{1}\Pi$          & $8.82$    ($0.072$)  &  $8.36$    ($0.093$)   &       $8.72$       &  $9.55$   & $9.03$       ($0.140$)   & $8.42$        ($0.134$)   &  $8.60$  \\
			&       $(1\pi)^{3}(5\si)^{2}(2\pi)^{1}$ & ${}^{3}\Delta$       &        $9.46$        &         $9.27$         &       $9.40$       &  $9.57$   &          $9.84$          &          $9.71$           &  $9.44$  \\
			&       $(1\pi)^{3}(5\si)^{2}(2\pi)^{1}$ & ${}^{3}\Sigma^{-}$   &       $10.15$        &         $9.77$         &       $9.98$       &  $10.16$  &         $10.47$          &          $9.68$           &  $9.95$  \\
			&       $(1\pi)^{3}(5\si)^{2}(2\pi)^{1}$ & ${}^{1}\Sigma^{-}$   &       $10.21$        &        $10.02$         &      $10.21$       &  $10.32$  &         $10.58$          &          $10.43$          & $10.17$  \\
			&       $(1\pi)^{3}(5\si)^{2}(2\pi)^{1}$ & ${}^{1}\Delta$       &       $10.47$        &        $10.10$         &      $10.32$       &  $10.52$  &         $10.89$          &          $10.71$          & $10.28$  \\
			\hline
			\ce{N2}  & $(3\si_g)^{2}(1\pi_u)^{3}(1\pi_g)^{1}$ & ${}^{3}\Sigma_u^{+}$ &        $8.20$        &         $7.36$         &       $7.74$       &  $7.86$   &          $7.93$          &          $7.86$           &  $7.74$  \\
			& $(3\si_g)^{1}(1\pi_u)^{4}(1\pi_g)^{1}$ & ${}^{3}\Pi_g$        &        $8.25$        &         $7.85$         &       $8.15$       &  $8.14$   &          $8.82$          &          $8.38$           &  $8.09$  \\
			& $(3\si_g)^{1}(1\pi_u)^{4}(1\pi_g)^{1}$ & ${}^{3}\Delta_u$     &        $9.37$        &         $8.57$         &       $9.09$       &  $9.12$   &          $9.17$          &          $9.08$           &  $9.03$  \\
			& $(3\si_g)^{1}(1\pi_u)^{4}(1\pi_g)^{1}$ & ${}^{1}\Pi_g$        &        $9.60$        &         $9.32$         &       $9.53$       &  $9.41$   &         $10.04$          &          $9.59$           &  $9.45$  \\
			& $(3\si_g)^{2}(1\pi_u)^{3}(1\pi_g)^{1}$ & ${}^{3}\Sigma_u^{-}$ &       $10.36$        &         $9.36$         &      $10.00$       &  $10.16$  &         $10.29$          &          $10.14$          &  $9.82$  \\
			& $(3\si_g)^{2}(1\pi_u)^{3}(1\pi_g)^{1}$ & ${}^{1}\Sigma_u^{-}$ &       $10.43$        &         $9.62$         &      $10.25$       &  $10.22$  &         $10.23$          &          $10.14$          & $10.12$  \\
			& $(3\si_g)^{1}(1\pi_u)^{4}(1\pi_g)^{1}$ & ${}^{1}\Delta_u$     &       $10.95$        &        $10.00$         &      $10.66$       &  $10.81$  &         $10.90$          &          $10.77$          & $10.49$  \\
			\hline
			\ce{F2}  & $(1\pi_u)^{4}(1\pi_g)^{3}(3\si_u)^{1}$ & ${}^{3}\Pi_u$        &        $3.43$        &         $2.95$         &       $3.32$       &  $3.38$   &          $3.35$          &          $3.26$           &  $3.26$  \\
			& $(1\pi_u)^{4}(1\pi_g)^{3}(3\si_u)^{1}$ & ${}^{1}\Pi_u$        & $4.76$   ($<0.001$)  &  $4.18$    ($<0.001$)  &       $4.61$       &  $4.58$   &  $4.57$     ($<0.001$)   &  $4.49$       ($<0.001$)  &  $4.52$  \\
			& $(1\pi_u)^{3}(1\pi_g)^{4}(3\si_u)^{1}$ & ${}^{3}\Pi_g$        &        $7.28$        &         $6.50$         &       $7.05$       &  $7.07$   &          $6.96$          &          $6.81$           &  $6.84$  \\
			& $(1\pi_u)^{3}(1\pi_g)^{4}(3\si_u)^{1}$ & ${}^{3}\Sigma_u^{+}$ &        $7.42$        &         $6.62$         &       $7.07$       &  $7.41$   &          $7.35$          &          $7.27$           &  $7.12$  \\
			& $(1\pi_u)^{3}(1\pi_g)^{4}(3\si_u)^{1}$ & ${}^{1}\Pi_g$        &        $8.29$        &         $6.93$         &       $7.74$       &  $7.53$   &          $7.46$          &          $7.31$           &  $7.35$  \\
			& $(1\pi_u)^{4}(1\pi_g)^{2}(3\si_u)^{2}$ & ${}^{3}\Sigma_g^{-}$ &  \tnote{\textbf{a}}  &   \tnote{\textbf{a}}   & \tnote{\textbf{a}} &  $8.43$   &          $8.82$          &          $8.29$           &  $8.81$  \\
			\hline
			\ce{H2O} &       $(3a_1)^{2}(1b_1)^{1}(4a_1)^{1}$ & ${}^{3}B_1$          &        $6.63$        &         $7.40$         &       $7.04$       &  $7.11$   &          $7.40$          &          $7.17$           &  $7.14$  \\
			&       $(3a_1)^{2}(1b_1)^{1}(4a_1)^{1}$ & ${}^{1}B_1$          & $6.96$    ($0.055$)  &  $7.84$    ($0.059$)   &       $7.44$       &  $7.50$   &  $7.78$      ($0.053$)   &  $7.51$       ($0.052$)   &  $7.53$  \\
			&       $(3a_1)^{2}(1b_1)^{1}(2b_2)^{1}$ & ${}^{3}A_2$          &        $8.50$        &         $9.43$         &       $9.04$       &  $9.12$   &          $9.35$          &          $9.09$           &  $9.14$  \\
			&       $(3a_1)^{2}(1b_1)^{1}(2b_2)^{1}$ & ${}^{1}A_2$          &        $8.62$        &         $9.64$         &       $9.21$       &  $9.30$   &          $9.50$          &          $9.22$           &  $9.32$  \\
			&       $(3a_1)^{1}(1b_1)^{2}(4a_1)^{1}$ & ${}^{3}A_1$          &        $8.98$        &         $9.72$         &       $9.37$       &  $9.44$   &          $9.79$          &          $9.40$           &  $9.47$  \\
			&       $(3a_1)^{1}(1b_1)^{2}(4a_1)^{1}$ & ${}^{1}A_1$          & $9.35$     ($0.099$) & $10.26$    ($0.112$)   &       $9.85$       &  $9.95$   & $10.43$      ($0.120$)   &  $9.99$       ($0.112$)   &  $9.93$  \\
			&       $(3a_1)^{2}(1b_1)^{1}(2b_1)^{1}$ & ${}^{3}A_1$          &       $10.47$        &        $11.06$         &      $10.77$       &  $10.87$  &         $11.14$          &          $10.93$          & $10.91$  \\
			&       $(3a_1)^{2}(1b_1)^{1}(5a_1)^{1}$ & ${}^{3}B_1$          &       $10.44$        &        $11.31$         &      $10.93$       &  $11.00$  &         $11.33$          &          $11.08$          & $11.04$  \\
			&       $(3a_1)^{2}(1b_1)^{1}(5a_1)^{1}$ & ${}^{1}B_1$          & $10.59$  ($<0.001$)  &  $11.47$    ($0.003$)  &      $11.09$       &  $11.16$  & $11.57$      ($0.002$)   & $11.32$       ($0.002$)   & $11.19$  \\
			\hline
			\mae     &                                        &                      &       $0.478$        &        $0.339$         &      $0.120$       &  $0.126$  &         $0.301$          &          $0.139$          &          \\
			\std     &                                        &                      &       $0.510$        &        $0.362$         &      $0.137$       &  $0.190$  &         $0.162$          &          $0.179$          &          \\
			\maxe    &                                        &                      &       $0.942$        &        $0.529$         &      $0.392$       &  $0.949$  &         $0.728$          &          $0.521$          &          \\
			\hline
			\hline
		\end{tabular}
		\begin{tablenotes}
			\item[\textbf{a}] \scriptsize{Excited state with a highly-excited configuration, not present in single-reference calculations.}
		\end{tablenotes}
	\end{threeparttable}
\end{table*}

\begin{table*}[t!]
	\caption{Vertical excitation energies ($\Omega$, eV) and oscillator strengths ($f$) of molecules with stretched geometries. See \cref{sec:computational_details} for details of the calculations. Also shown are mean absolute errors (\mae), standard deviations (\std), and maximum absolute errors (\maxe) of the energies, relative to SHCI. Only orbitals participating in excitations are listed in the electron configuration of each excited state.}
	\label{tab:ee_2r_eq}
	\setlength{\extrarowheight}{2pt}
	\setstretch{1}
	\tiny
	\centering
	\hspace*{-1.2cm}
	\begin{threeparttable}
		\begin{tabular}{lrlccccccc}
			\hline\hline
			System   &                                        Configuration & Term                 &       ADC(2)        &        ADC(3)        &      EOM-CCSD      & sc-NEVPT2 &        MR-ADC(2)        &       MR-ADC(2)-X       &   SHCI   \\
			&                                                      &                      &   $\Omega \ (f)$    &    $\Omega \ (f)$    &      $\Omega$      & $\Omega$  &     $\Omega \ (f)$      &     $\Omega \ (f)$      & $\Omega$ \\
			\hline
			\ce{HF}  &                     $(1\pi)^{3}(3\si)^{2}(4\si)^{1}$ & ${}^{3}\Pi$          &       $1.12$        &        $1.38$        &       $1.80$       &  $1.49$   &         $1.60$          &         $1.55$          &  $1.59$  \\
			&                     $(1\pi)^{3}(3\si)^{2}(4\si)^{1}$ & ${}^{1}\Pi$          & $1.35$    ($0.001$) & $1.59$    ($<0.001$) &       $1.98$       &  $1.62$   & $1.74$      ($<0.001$)  & $1.69$      ($<0.001$)  &  $1.72$  \\
			&                     $(1\pi)^{4}(3\si)^{1}(4\si)^{1}$ & ${}^{3}\Sigma^{+}$   &       $1.38$        &        $1.14$        &       $1.51$       &  $1.72$   &         $1.74$          &         $1.72$          &  $1.77$  \\
			&                     $(1\pi)^{4}(3\si)^{1}(4\si)^{1}$ & ${}^{1}\Sigma^{+}$   & $5.98$    ($0.555$) & $4.14$    ($0.289$)  &       $6.48$       &  $6.38$   & $6.31$       ($0.343$)  & $6.24$       ($0.335$)  &  $6.03$  \\
			&                     $(1\pi)^{3}(3\si)^{2}(5\si)^{1}$ & ${}^{3}\Pi$          &       $7.68$        &       $13.32$        &      $10.73$       &  $10.40$  &         $11.25$         &         $10.16$         & $10.50$  \\
			&                     $(1\pi)^{3}(3\si)^{2}(5\si)^{1}$ & ${}^{1}\Pi$          & $7.68$    ($0.019$) & $13.13$    ($0.008$) &      $10.80$       &  $10.49$  & $11.37$      ($0.022$)  & $10.22$       ($0.015$) & $10.59$  \\
			&                     $(1\pi)^{4}(3\si)^{1}(5\si)^{1}$ & ${}^{3}\Sigma^{+}$   &       $10.67$       &       $11.46$        &      $11.57$       &  $11.58$  &         $11.66$         &         $11.51$         & $11.62$  \\
			&                     $(1\pi)^{4}(3\si)^{1}(5\si)^{1}$ & ${}^{1}\Sigma^{+}$   & $10.85$   ($0.074$) & $11.52$   ($0.011$)  &      $11.65$       &  $11.61$  & $11.73$      ($<0.001$) & $11.56$       ($0.001$) & $11.66$  \\
			\hline
			\ce{CO}  &                     $(5\si)^{2}(1\pi)^{3}(2\pi)^{1}$ & ${}^{3}\Sigma^{+}$   &       $2.57$        &       $-3.43$        &      $-1.74$       &  $0.42$   &         $0.43$          &         $0.42$          &  $0.42$  \\
			&                     $(5\si)^{2}(1\pi)^{3}(2\pi)^{1}$ & ${}^{1}\Delta$       &       $2.49$        &       $-3.25$        &      $-1.95$       &  $0.77$   &         $0.79$          &         $0.77$          &  $0.76$  \\
			&                     $(5\si)^{2}(1\pi)^{3}(2\pi)^{1}$ & ${}^{1}\Sigma^{-}$   &       $2.53$        &       $-3.23$        &      $-1.89$       &  $0.78$   &         $0.80$          &         $0.78$          &  $0.77$  \\
			&                     $(5\si)^{2}(1\pi)^{3}(2\pi)^{1}$ & ${}^{3}\Delta$       &       $2.52$        &       $-3.45$        &      $-1.71$       &  $0.81$   &         $0.83$          &         $0.81$          &  $0.80$  \\
			&                     $(5\si)^{2}(1\pi)^{3}(2\pi)^{1}$ & ${}^{3}\Sigma^{-}$   &       $2.53$        &       $-3.42$        &      $-1.71$       &  $0.90$   &         $0.93$          &         $0.90$          &  $0.88$  \\
			&                     $(1\pi)^{2}(2\pi)^{2}(6\si)^{2}$ & ${}^{5}\Sigma^{+}$   & \tnote{\textbf{a}}  &  \tnote{\textbf{a}}  & \tnote{\textbf{a}} &  $1.28$   &         $1.44$          &         $1.41$          &  $1.35$  \\
			&                     $(1\pi)^{3}(2\pi)^{2}(6\si)^{1}$ & ${}^{3}\Pi$          & \tnote{\textbf{a}}  &  \tnote{\textbf{a}}  & \tnote{\textbf{a}} &  $1.44$   &         $1.52$          &         $1.45$          &  $1.44$  \\
			\hline
			\ce{N2}  &               $(3\si_g)^{2}(1\pi_u)^{3}(1\pi_g)^{1}$ & ${}^{3}\Sigma_u^{+}$ &       $6.26$        &       $-22.30$       &      $-2.92$       &  $0.15$   &         $0.15$          &         $0.14$          &  $0.15$  \\
			&               $(3\si_g)^{2}(1\pi_u)^{2}(1\pi_g)^{2}$ & ${}^{5}\Sigma_g^{+}$ & \tnote{\textbf{a}}  &  \tnote{\textbf{a}}  & \tnote{\textbf{a}} &  $0.46$   &         $0.54$          &         $0.53$          &  $0.47$  \\
			&   $(3\si_g)^{1}(1\pi_u)^{2}(1\pi_g)^{2}(3\si_u)^{1}$ & ${}^{7}\Sigma_u^{+}$ & \tnote{\textbf{a}}  &  \tnote{\textbf{a}}  & \tnote{\textbf{a}} &  $1.10$   &         $1.43$          &         $1.43$          &  $1.09$  \\
			&               $(3\si_g)^{2}(1\pi_u)^{3}(1\pi_g)^{1}$ & ${}^{3}\Delta_u$     &       $8.08$        &       $-20.88$       &       $3.45$       &  $2.40$   &         $2.41$          &         $2.37$          &  $2.46$  \\
			&               $(3\si_g)^{2}(1\pi_u)^{2}(1\pi_g)^{2}$ & ${}^{3}\Delta_g$     & \tnote{\textbf{a}}  &  \tnote{\textbf{a}}  & \tnote{\textbf{a}} &  $2.60$   &         $2.62$          &         $2.58$          &  $2.72$  \\
			&               $(3\si_g)^{1}(1\pi_u)^{4}(1\pi_g)^{1}$ & ${}^{3}\Pi_g$        &       $7.80$        &       $-19.35$       &       $3.72$       &  $2.97$   &         $2.98$          &         $2.93$          &  $2.95$  \\
			&               $(3\si_g)^{1}(1\pi_u)^{4}(3\si_u)^{1}$ & ${}^{3}\Sigma_u^{+}$ &       $8.12$        &       $-16.99$       &       $3.83$       &  $3.22$   &         $3.24$          &         $3.21$          &  $3.24$  \\
			\hline
			\ce{F2}  &   $(1\pi_u)^{4}(1\pi_g)^{3}(3\si_g)^{2}(3\si_u)^{1}$ & ${}^{3}\Pi_g$        &       $10.37$       &       $-4.60$        &       $0.14$       &  $0.01$   &         $0.10$          &         $-0.17$         & $-0.02$  \\
			&   $(1\pi_u)^{3}(1\pi_g)^{4}(3\si_g)^{2}(3\si_u)^{1}$ & ${}^{3}\Pi_u$        &       $1.24$        &       $-4.55$        &       $0.06$       &  $0.05$   &         $0.12$          &         $-0.16$         &  $0.01$  \\
			&   $(1\pi_u)^{3}(1\pi_g)^{4}(3\si_g)^{2}(3\si_u)^{1}$ & ${}^{1}\Pi_u$        &       $1.45$        &       $-4.29$        &       $0.17$       &  $0.03$   &         $0.07$          &         $-0.13$         &  $0.01$  \\
			&   $(1\pi_u)^{4}(1\pi_g)^{3}(3\si_g)^{2}(3\si_u)^{1}$ & ${}^{1}\Pi_g$        &       $10.45$       &       $-4.35$        &       $0.19$       &  $0.04$   &         $0.06$          &         $-0.16$         &  $0.02$  \\
			&   $(1\pi_u)^{4}(1\pi_g)^{4}(3\si_g)^{1}(3\si_u)^{1}$ & ${}^{3}\Sigma_u^{+}$ &       $11.90$       &       $-8.31$        &      $-1.31$       &  $0.03$   &         $0.03$          &         $-0.02$         &  $0.03$  \\
			\hline
			\ce{H2O} &           $(1b_1)^{1}(1b_2)^{2}(3a_1)^{2}(4a_1)^{1}$ & ${}^{3}B_1$          &       $0.73$        &       $-0.32$        &       $1.12$       &  $0.88$   &         $0.94$          &         $0.89$          &  $0.87$  \\
			&           $(1b_1)^{1}(1b_2)^{2}(3a_1)^{2}(4a_1)^{1}$ & ${}^{1}B_1$          &  $1.23$  ($0.001$)  &  $-0.22$  ($<0.001$) &       $1.33$       &  $1.02$   & $1.08$       ($<0.001$) &  $1.05$      ($<0.001$) &  $1.04$  \\
			&           $(1b_1)^{2}(1b_2)^{2}(3a_1)^{1}(4a_1)^{1}$ & ${}^{3}A_1$          &       $1.12$        &       $-0.13$        &       $0.50$       &  $1.08$   &         $1.09$          &         $1.07$          &  $1.10$  \\
			&           $(1b_1)^{1}(1b_2)^{2}(3a_1)^{2}(2b_2)^{1}$ & ${}^{3}A_2$          &       $1.46$        &       $-0.23$        &       $1.70$       &  $1.24$   &         $1.30$          &         $1.25$          &  $1.24$  \\
			&           $(1b_1)^{2}(1b_2)^{1}(3a_1)^{2}(4a_1)^{1}$ & ${}^{3}B_2$          &       $1.78$        &        $0.94$        &       $1.84$       &  $1.52$   &         $1.54$          &         $1.51$          &  $1.54$  \\
			&           $(1b_1)^{1}(1b_2)^{2}(3a_1)^{2}(2b_2)^{1}$ & ${}^{1}A_2$          &       $1.87$        &        $0.37$        &       $1.88$       &  $1.54$   &         $1.58$          &         $1.54$          &  $1.55$  \\
			& $(1b_1)^{1}(1b_2)^{2}(3a_1)^{1}(4a_1)^{1}(2b_2)^{1}$ & ${}^{5}A_2$          & \tnote{\textbf{a}}  &  \tnote{\textbf{a}}  & \tnote{\textbf{a}} &  $2.14$   &         $2.39$          &         $2.33$          &  $2.16$  \\
			& $(1b_1)^{1}(1b_2)^{1}(3a_1)^{2}(4a_1)^{1}(2b_2)^{1}$ & ${}^{5}B_1$          & \tnote{\textbf{a}}  &  \tnote{\textbf{a}}  & \tnote{\textbf{a}} &  $2.25$   &         $2.50$          &         $2.44$          &  $2.28$  \\
			&           $(1b_1)^{1}(1b_2)^{2}(3a_1)^{1}(4a_1)^{2}$ & ${}^{3}B_1$          & \tnote{\textbf{a}}  &  \tnote{\textbf{a}}  & \tnote{\textbf{a}} &  $2.34$   &         $2.54$          &         $2.45$          &  $2.38$  \\
			\hline
			\mae     &                                                      &                      &       $2.705$       &       $5.368$        &      $0.850$       &  $0.042$  &         $0.114$         &         $0.091$         &          \\
			\std     &                                                      &                      &       $3.815$       &       $7.499$        &      $1.239$       &  $0.074$  &         $0.187$         &         $0.133$         &          \\
			\maxe    &                                                      &                      &      $11.870$       &       $23.338$       &      $3.068$       &  $0.348$  &         $0.782$         &         $0.367$         &          \\ \hline\hline
		\end{tabular}
		\begin{tablenotes}
			\item[\textbf{a}] \scriptsize{Excited state with a highly-excited configuration, not present in single-reference calculations.}
		\end{tablenotes}
	\end{threeparttable}
\end{table*}

The active spaces used in the MR-ADC and sc-NEVPT2 computations (\cref{sec:results:diatomic_water}) were chosen as follows:
\begin{align*}
	\ce{HF}:   &\ (3\si)^{2} (1\pi)^{4} \\
	           &\ (4\si)^{0} (5\si)^{0} (2\pi)^{0} (3\pi)^{0} \\
	\ce{CO}:   &\ (1\pi)^{4} (5\si)^{2} \\
	           &\ (2\pi)^{0} (6\si)^{0} (7\si)^{0} (3\pi)^{0} \\
	\ce{N2}:   &\ (3\si_g)^{2} (1\pi_u)^{4} \\
	           &\ (3\si_u)^{0} (1\pi_g)^{0} (4\si_g)^{0} (2\pi_u)^{0} \\
	\ce{F2}:   &\ (1\pi_u)^{4} (3\si_g)^{2} (1\pi_g)^{4} \\
	           &\ (3\si_u)^{0} (2\pi_u)^{0} \\
	\ce{H2O}:  &\ (1b_2)^{2} (3a_1)^{2} (1b_1)^{2} \\
	           &\ (4a_1)^{0} (2b_2)^{0} (5a_1)^{0} (2b_1)^{0} (6a_1)^{0} (3b_2)^{0} \\
	\ce{C2}:   &\ (2\si_g)^{2} (2\si_u)^{2} (1\pi_u)^{4} \\
	           &\ (3\si_g)^{0} (1\pi_g)^{0} (3\si_u)^{0}
\end{align*}
where the molecular orbital occupations shown correspond to the those in the Hartree--Fock Slater determinant. The active spaces used in the calculations of the alkene molecules ranged from the full valence $\pi$-orbital space up to the triple-$\pi$ space for ethylene and up to double-$\pi$ for butadiene, as described in the Supporting Information. The sc-NEVPT2 computations were performed using the state-averaged CASSCF reference wavefunctions, where the ground and all excited states of interest were averaged with equal weights. All MR-ADC calculations were performed using the state-specific (ground-state) CASSCF reference wavefunctions. The MR-ADC calculations were converged with the number of CASCI states by including at least 15 excited singlet CASCI states and no less than 10 excited triplet CASCI states. The $\eta_{s}$ = $10^{-5}$ and $\eta_{d}$ = $10^{-10}$ truncation parameters were used to eliminate redundant excitations in the solution of the MR-ADC equations.\cite{Chatterjee:2019p5908,Chatterjee:2020p6343} As in our previous work,\cite{Chatterjee:2019p5908,Chatterjee:2020p6343} we have confirmed that the computed MR-ADC excitation energies and oscillator strengths are size-consistent (see the Supporting Information for details).

\section{Results}
\label{sec:results}

\subsection{Excited States in \ce{HF}, \ce{CO}, \ce{N2}, \ce{F2}, and \ce{H2O}}
\label{sec:results:diatomic_water}

We begin our discussion of results by benchmarking the MR-ADC(2) and MR-ADC(2)-X excitation energies against the accurate excited-state energies computed using the semi-stochastic heat-bath CI algorithm (SHCI)\cite{Holmes:2016p3674,Sharma:2017p1595,Holmes:2017p164111} for five small molecules (\ce{HF}, \ce{CO}, \ce{N2}, \ce{H2O}, and \ce{F2}) at equilibrium and stretched geometries. In addition to SHCI, we compare the performance of our MR-ADC methods with the conventional (single-reference) ADC approximations (ADC(n), $n = 2,3$), equation-of-motion coupled cluster theory with single and double excitations (EOM-CCSD), as well as strongly-contracted second-order N-electron valence perturbation theory (sc-NEVPT2).

\cref{tab:ee_r_eq} compares vertical excitation energies of molecules at their near-equilibrium geometries computed using six approximate methods with the reference results from SHCI. EOM-CCSD shows the best performance for this benchmark set, with the mean absolute error \mae = 0.12 eV and a standard deviation (\std) of 0.14 eV. The single-reference ADC(2) method shows the largest \mae = 0.48 eV and \std = 0.51 eV, while ADC(3) exhibits only a modest improvement lowering those values to 0.34 eV (\mae) and 0.36 eV (\std). MR-ADC(2) outperforms ADC(3) with smaller \mae = 0.30 eV and \std = 0.16 eV, due to the high-order description of electron correlation effects in the active space. Incorporating the third-order terms in MR-ADC(2)-X reduces \mae by a factor of two (\mae = 0.14 eV, \std = 0.18 eV). While the MR-ADC(2)-X \mae and \std are quite similar to those of sc-NEVPT2 (\mae = 0.13 eV, \std = 0.19 eV), its maximum error (\maxe = 0.52 eV) is almost a factor of two smaller (\maxe = 0.95 eV for sc-NEVPT2). \cref{tab:ee_r_eq} also reports the oscillator strengths ($f$) computed using the single- and multireference ADC methods. The $f$ values computed with different approximations generally agree well with each other, with a couple of exceptions for the ${}^{1}\Pi$ state of \ce{HF} and ${}^{1}\Pi$ state of \ce{CO} where the computed oscillator strengths vary by up to a factor of two.

\begin{figure*}[t!]
    \subfloat[]{\label{fig:ee_mae_std_re}\includegraphics[width=0.45\textwidth]{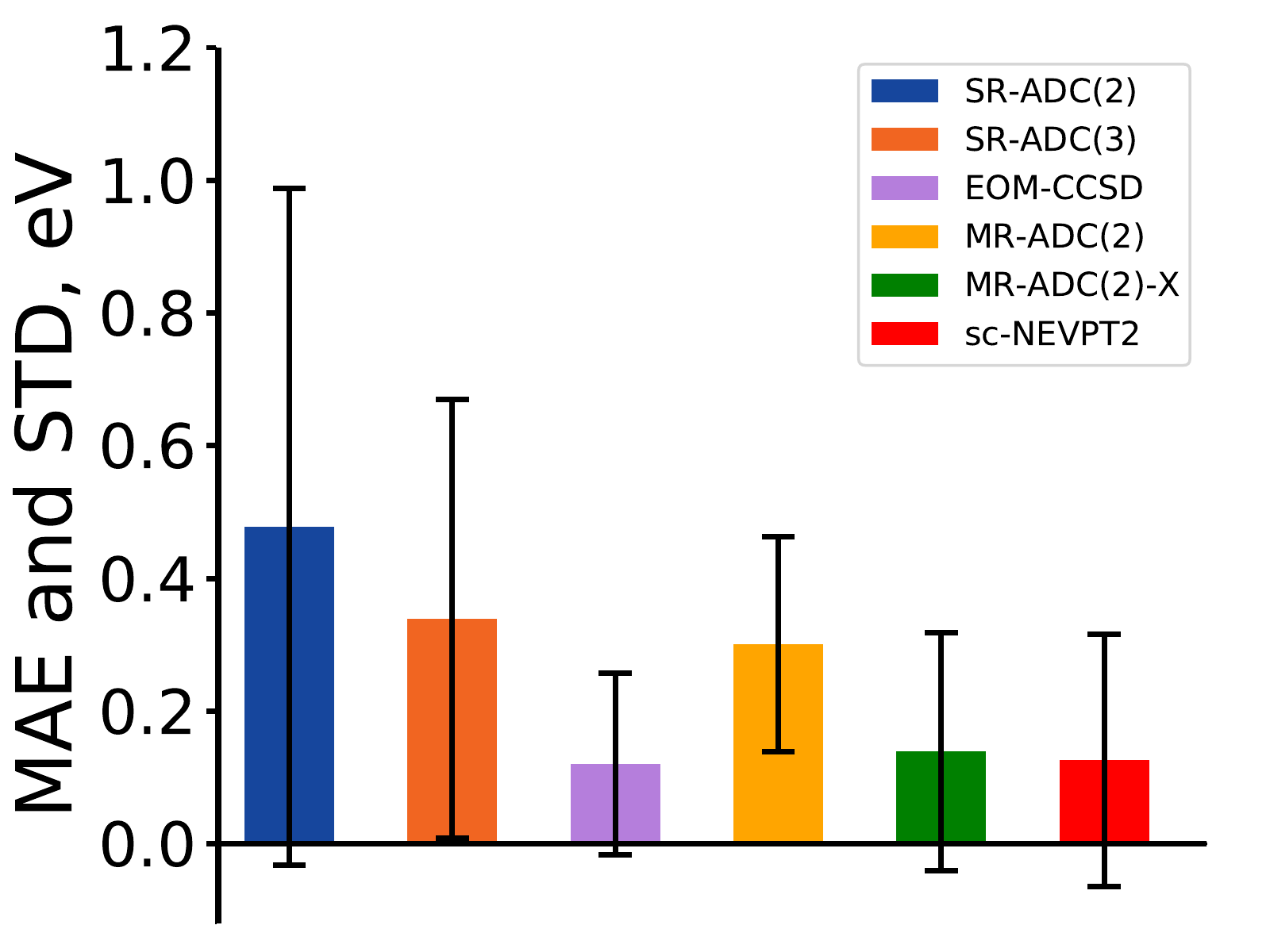}} \quad
    \subfloat[]{\label{fig:ee_mae_std_2re}\includegraphics[width=0.45\textwidth]{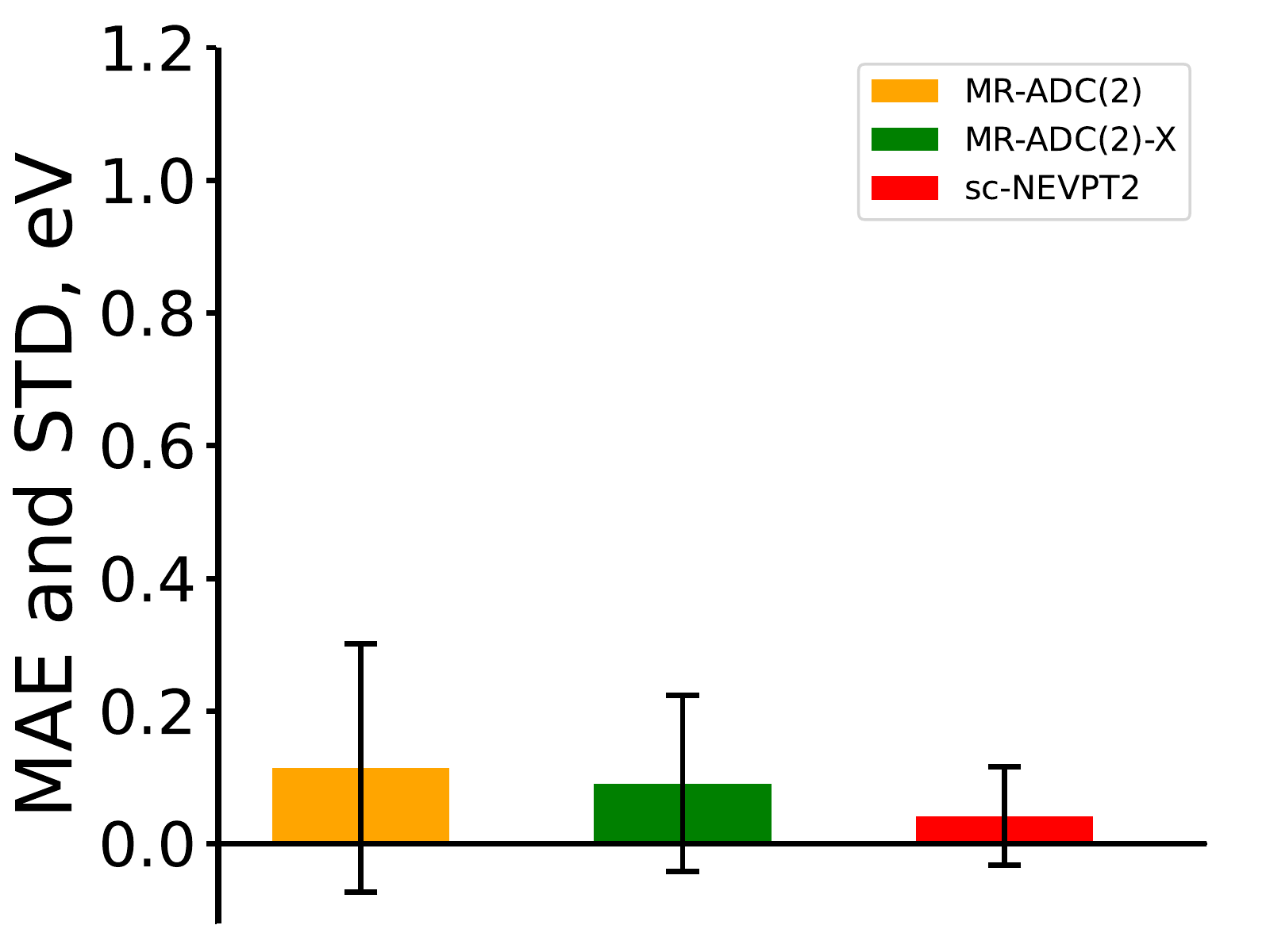}}
    \captionsetup{justification=raggedright,singlelinecheck=false}
    \caption{Mean absolute errors (MAE, eV) and standard deviations from the mean signed error (STD, eV) for vertical electron excitation energies of molecules with (a) equilibrium and (b) stretched geometries, relative to SHCI. MAE is represented as a height of each colored bar, while STD is depicted as a radius of the black vertical line. See \cref{tab:ee_r_eq,tab:ee_2r_eq} for data on individual molecules with all methods.}
    \label{fig:ee_mae_std}
\end{figure*}

\cref{tab:ee_2r_eq} shows results for electronic excitations of molecules with stretched geometries where the description of multireference effects is important. Unsurprisingly, the accuracy of results produced by the single-reference methods decreases dramatically. In particular, the errors of the single-reference ADC(n) methods become larger with increasing perturbation order (n) as \mae roughly doubles from ADC(2) (2.71 eV) to ADC(3) (5.37 eV), while \mae for the EOM-CCSD method increases eightfold. The best agreement with SHCI is demonstrated by sc-NEVPT2 (\mae = 0.04 eV, \std = 0.07 eV). The MR-ADC methods perform similarly well with \mae of 0.11 and 0.09 eV for MR-ADC(2) and MR-ADC(2)-X, respectively. The MR-ADC(2)-X approximation shows a smaller \maxe error (0.37 eV) than that of MR-ADC(2) (0.78 eV), close to \maxe of sc-NEVPT2 (0.35 eV).

The performance of all methods for this benchmark set is summarized in \cref{fig:ee_mae_std}. Overall, our results demonstrate that MR-ADC(2) and MR-ADC(2)-X produce accurate results for many electronic transitions outperforming the single-reference ADC(2) and ADC(3) methods at near-equilibrium geometries and showing accuracy similar to sc-NEVPT2 near dissociation. In contrast to sc-NEVPT2, the MR-ADC methods allow for a straightforward calculation of transition properties (such as oscillator strengths) and do not require using state-averaged CASSCF reference wavefunctions. 

\subsection{Carbon Dimer}
\label{sec:results:carbon_dimer}

\begin{table*}[t!]
	\caption{Carbon dimer vertical excitation energies ($\Omega$, eV) and oscillator strengths ($f$) computed using the cc-pVDZ basis set with $r$\ce{(C-C)} = 2.4 \bohr. The MR-ADC and sc-NEVPT2 calculations were performed using the state-specific and state-averaged CASSCF reference wavefunctions, respectively, with the same $(8e, 8o)$ active space. Also shown are mean absolute errors (\mae), standard deviations (\std), and maximum absolute errors (\maxe) in the computed excitation energies, relative to the DMRG results from Ref.\@ \citenum{Wouters:2014p1501}.}
	\label{tab:ee_c2_dimer}
	\setlength{\extrarowheight}{2pt}
	\setstretch{1}
	\footnotesize
	\centering
	\begin{threeparttable}
		\begin{tabular}{lrlcccc}
			\hline
			\hline
			System  & Configuration                                          & Term                 & sc-NEVPT2 &       MR-ADC(2)        &       MR-ADC(2)-X       &   DMRG   \\
			        &                                                        &                      & $\Omega$  &   $\Omega \ (f) $   &      $\Omega \ (f)$       & $\Omega$ \\
			        \hline
			\ce{C2} & $(2\si_u)^{2} (1\pi_u)^{3} (3\si_g)^{1}$               & ${}^{3}\Pi_u$        &  $0.24$   &         $0.23$         &         $0.19$          &  $0.21$  \\
			        & $(2\si_u)^{2} (1\pi_u)^{3} (3\si_g)^{1}$               & ${}^{1}\Pi_u$        &  $1.36$   & $1.31$      ($0.003$)  & $1.28$       ($0.003$)  &  $1.29$  \\
			        & $(2\si_u)^{2} (1\pi_u)^{2} (3\si_g)^{2}$               & ${}^{3}\Sigma_g^{-}$ &  $1.33$   &         $1.27$         &         $1.25$          &  $1.29$  \\
			        & $(2\si_u)^{1} (1\pi_u)^{4} (3\si_g)^{1}$               & ${}^{3}\Sigma_u^{+}$ &  $1.33$   &         $1.53$         &         $1.38$          &  $1.29$  \\
			        & $(2\si_u)^{2} (1\pi_u)^{2} (3\si_g)^{2}$               & ${}^{1}\Delta_g$     &  $2.22$   &         $2.38$         &         $2.24$          &  $2.17$  \\
			        & $(2\si_u)^{2} (1\pi_u)^{2} (3\si_g)^{2}$               & ${}^{1}\Sigma_g^{+}$ &  $2.49$   &         $2.64$         &         $2.51$          &  $2.45$  \\
			        & $(2\si_u)^{1} (1\pi_u)^{3} (3\si_g)^{2}$               & ${}^{3}\Pi_g$        &  $2.71$   &         $2.74$         &         $2.61$          &  $2.65$  \\
			        & $(2\si_u)^{1} (1\pi_u)^{3} (3\si_g)^{2}$               & ${}^{1}\Pi_g$        &  $4.72$   &         $4.71$         &         $4.61$          &  $4.61$  \\
			        & $(2\si_u)^{2} (1\pi_u)^{3} (1\pi_g)^{1}$ 		         & ${}^{3}\Delta_u$     &  $6.96$   &         $6.79$         &         $6.60$          &  $6.66$  \\
			        \hline
			\mae    &                                                        &                      &  $0.082$  &        $0.114$         &         $0.045$         &          \\
			\std    &                                                        &                      &  $0.084$  &        $0.093$         &         $0.055$         &          \\
			\maxe   &                                                        &                      &  $0.297$  &        $0.239$         &         $0.086$         &          \\
			\hline
			\hline
		\end{tabular}
	\end{threeparttable}
\end{table*}

Next, we consider the \ce{C2} molecule, known for the multireference nature of its ground and excited states, which also require accurate description of dynamic correlation.\cite{Roos:1987p399,Bauschlicher:1987p1919,Watts:1992p6073,Abrams:2004p9211,Wouters:2014p1501,Holmes:2017p164111} 
In our earlier work,\cite{Sokolov:2018p204113} we demonstrated that the first-order MR-ADC approximation (MR-ADC(1)) captures the strongly correlated nature of the \ce{C2} excited states, but produces large errors in vertical excitation energies due to the missing description of the two-electron dynamic correlation effects. Here, we reinvestigate \ce{C2} using MR-ADC(2) and MR-ADC(2)-X that provide a more balanced description of static and dynamic correlation.

\begin{figure*}[t!]
	\centering
	\subfloat[]{\label{fig:c2_states_barchart}\includegraphics[width=0.45\textwidth]{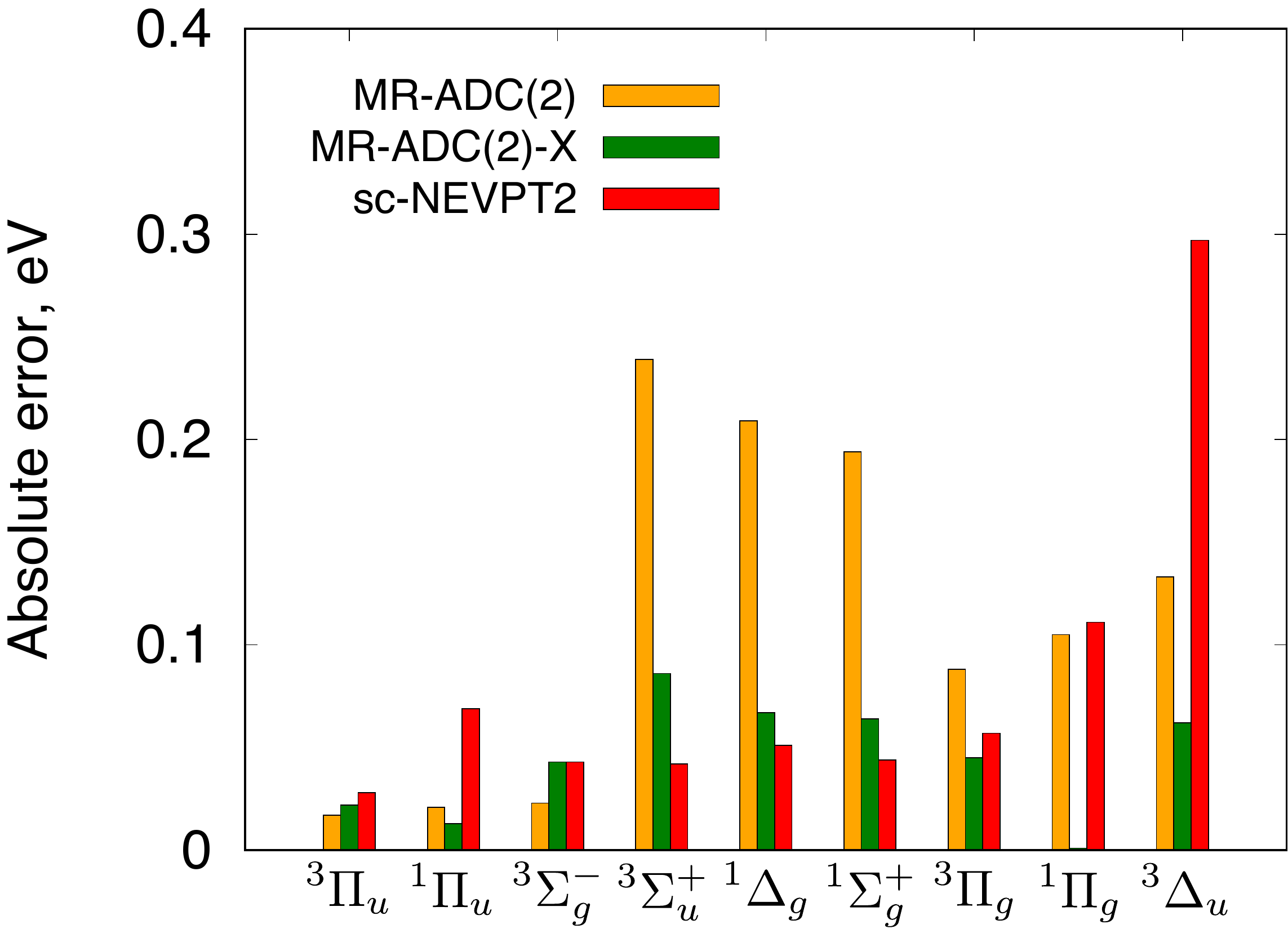}} \quad
	\subfloat[]{\label{fig:c2_barchart}\includegraphics[width=0.43\textwidth]{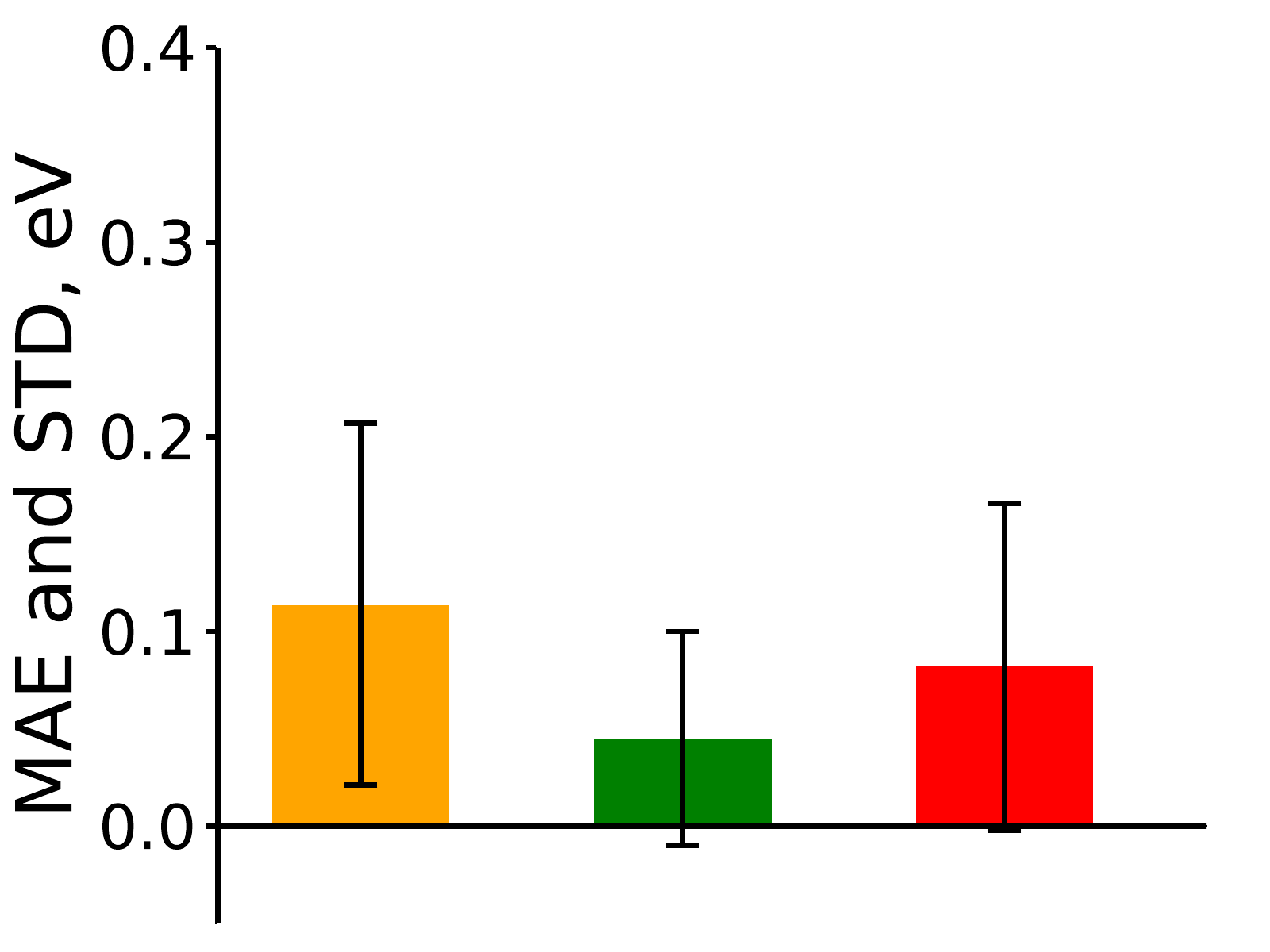}}
	\captionsetup{justification=raggedright,singlelinecheck=false}
	\caption{Absolute errors for vertical excitation energies of \ce{C2} (plot (a)), as well as mean absolute errors (MAE) and standard deviations (STD) (plot (b)), relative to DMRG.\cite{Wouters:2014p1501} In plot (b), the MAE is represented as the height of each colored bar, while STD is depicted as a radius of the black vertical line. See \cref{tab:ee_c2_dimer} for vertical excitation energies. }
	\label{fig:ee_mae_std_c2_states}
\end{figure*}

\cref{tab:ee_c2_dimer} reports the MR-ADC(2), MR-ADC(2)-X, and sc-NEVPT2 vertical excitation energies for \ce{C2} at a near equilibrium bond distance (\mbox{$r$ = 2.4 $a_0$}) computed using the cc-pVDZ basis set and $(8e, 8o)$ active space. The results of these multireference methods are compared to the accurate excitation energies from density matrix renormalization group (DMRG) computed by Wouters et al.\cite{Wouters:2014p1501} \cref{fig:ee_mae_std_c2_states} summarizes the performance of each method for predicting the vertical excitation energies of \ce{C2}. The best agreement with DMRG is shown by MR-ADC(2)-X that reproduces the reference energies with \mae and \std of $\sim$ 0.05 eV. The sc-NEVPT2 method (\mae and \std of 0.08 eV) exhibits significantly worse accuracy than MR-ADC(2)-X, but performs slightly better than MR-ADC(2) (\mae = 0.11 eV, \std = 0.09). The MR-ADC(2) and MR-ADC(2)-X mean absolute errors are reduced tenfold compared to that of MR-ADC(1),\cite{Sokolov:2018p204113} indicating that the second-order contributions to the MR-ADC propagator are absolutely critical for quantitive predictions of the \ce{C2} excited-state energies. As illustrated in \cref{fig:c2_states_barchart}, MR-ADC(2)-X has a fairly consistent performance for all excited states resulting in a small \std, while MR-ADC(2) shows significantly larger errors ($\sim$ 0.2 eV) for the ${}^{3}\Sigma_u^{+}$, ${}^{1}\Delta_g$, and ${}^{1}\Sigma_g^{+}$ states. The errors of sc-NEVPT2 gradually increase reaching $\sim$ 0.3 eV for the highest-energy ${}^{3}\Delta_u$ state. The MR-ADC methods predict only one dipole-allowed transition (${}^{1}\Pi_u$) with an oscillator strength of 0.003, in agreement with the selection rules of optical spectroscopy. 

\subsection{Alkenes: Ethylene and Butadiene}
\label{sec:results:alkenes}

Finally, we benchmark MR-ADC(2) and MR-ADC(2)-X for the low-lying excited states of ethylene (\ce{C2H4}) and butadiene (\ce{C4H6}). Both molecules have a low-lying triplet state (${}^{3}B_{1u}$ or ${}^{3}B_{u}$), as well as a dipole-allowed $\pi-\pi^*$ transition to the singlet state (${}^{1}B_{1u}$ or ${}^{1}B_{u}$)  that requires a very accurate description of dynamic correlation between the $\sigma$ and $\pi$ electrons.\cite{Davidson:1996p6161,Angeli:2010p2436,Watts:1998p6979,Copan:2018p4097,Chien:2018p2714} In addition, the butadiene molecule features a dipole-forbidden transition to the ${}^{1}A_{g}$ state with a substantial double-excitation character, requiring an accurate description of static correlation in the $\pi$ and $\pi^{*}$ orbitals. For this reason, excitation energies of the butadiene ${}^{1}B_{u}$ and ${}^{1}A_{g}$ states are very sensitive to the level of electron correlation treatment and have been a subject of many theoretical studies.\cite{Watts:1998p6979,Copan:2018p4097,Chien:2018p2714,Tavan:1986p6602,Tavan:1987p4337,Nakayama:1998p157,Li:1999p177,Kurashige:2004p425,Starcke:2006p39,Ghosh:2008p144117,Schreiber:2008p134110,Zgid:2009p194107,Daday:2012p4441,Watson:2012p4013,Sokolov:2017p244102,Zimmerman:2017p4712}. In this work, we compute the ethylene and butadiene excited states using MR-ADC(2) and MR-ADC(2)-X and compare their excitation energies to accurate reference results from SHCI obtained by Chien et al.\cite{Chien:2018p2714} for the same molecular geometries and basis set (see \cref{sec:computational_details} for details). 

\begin{table}[t!]
	\caption{Vertical excitation energies (eV) of the ethylene molecule (\ce{C2H4}) for the lowest-energy triplet state (${}^{3}B_{1u}$) and the lowest-energy dipole-allowed singlet state (${}^{1}B_{1u}$). The MR-ADC results also show the oscillator strength strengths ($f$) in parentheses. See \cref{sec:computational_details} for details of the calculations.}
		\label{tab:ee_ethyl}
		\setlength{\extrarowheight}{2pt}
		\setstretch{1}
		\footnotesize
		\centering
		\begin{threeparttable}
			\begin{tabular}{ccccc}
				\hline\hline
				          Method           & Active Space &     \multicolumn{2}{c}{States}     &  \\
				                           &              & ${}^{3}B_{1u}$ &  ${}^{1}B_{1u}$  &  \\ \hline
				EOM-CCSD\tnote{\textbf{a}} &              &     $4.46$     &      $8.14$      &  \\ \hline
				        sc-NEVPT2          &  $(2e, 2o)$   &     $4.84$     &      $7.68$      &  \\
				                           &  $(2e, 4o)$   &     $4.64$     &      $8.28$      &  \\
				                           &  $(2e, 6o)$   &     $4.60$     &      $8.34$      &  \\ \hline
				        MR-ADC(2)          &  $(2e, 2o)$   &     $4.85$     & $8.19$ ($0.365$) &  \\
				                           &  $(2e, 4o)$   &     $4.71$     & $8.17$ ($0.363$) &  \\
				                           &  $(2e, 6o)$   &     $4.66$     & $8.27$ ($0.363$) &  \\ \hline
				       MR-ADC(2)-X         &  $(2e, 2o)$   &     $4.68$     & $7.82$ ($0.338$) &  \\
				                           &  $(2e, 4o)$   &     $4.56$     & $7.92$ ($0.343$) &  \\
				                           &  $(2e, 6o)$   &     $4.51$     & $8.03$ ($0.344$) &  \\ \hline
				  SHCI\tnote{\textbf{b}}   &              &     $4.59$     &      $8.05$      &  \\ \hline\hline
			\end{tabular}
				\begin{tablenotes}
					\item[\textbf{a}] 	\footnotesize{From Ref.\@ \citenum{Copan:2018p4097}}
					\item[\textbf{b}] 	\footnotesize{From Ref.\@ \citenum{Chien:2018p2714}}
				\end{tablenotes}
		\end{threeparttable}
\end{table}

\cref{tab:ee_ethyl} reports vertical excitation energies for the ${}^{3}B_{u}$ and ${}^{1}B_{u}$ excited states of ethylene computed using MR-ADC(2), MR-ADC(2)-X, sc-NEVPT2, and EOM-CCSD, along with the reference results from SHCI. For the multireference methods we employ three different active spaces ranging from the full-$\pi$ $(2e, 2o)$ to the triple-$\pi$ $(2e, 6o)$ active space (see the Supporting Information for details). The EOM-CCSD excitation energies are in a good agreement with SHCI with errors less than 0.15 eV, indicating that a high-level description of dynamic correlation is sufficient for accurate description of the ${}^{3}B_{u}$ and ${}^{1}B_{u}$ excited states. Using the minimal $(2e, 2o)$ active space, all three multireference methods overestimate the relative energy of the ${}^{3}B_{u}$ state with errors ranging from 0.09 eV (MR-ADC(2)-X) to 0.26 eV (MR-ADC(2)). Increasing the active space lowers the computed ${}^{3}B_{u}$ excitation energy reducing the errors of all three methods to less than 0.1 eV. In contrast, including more $\pi$-orbitals in the active space increases the computed energies of the ${}^{1}B_{u}$ state. Out of all multireference methods, sc-NEVPT2 shows the strongest active-space dependence of the ${}^{1}B_{u}$ excitation energy that changes from 7.68 $(2e, 2o)$ to 8.34 eV $(2e, 6o)$ and has the worst agreement with SHCI for the largest active space with 0.29 eV error. MR-ADC(2)-X combined with the $(2e, 6o)$ active space shows the best agreement with SHCI for the ${}^{1}B_{u}$ state with a small 0.02 eV error. The performance of MR-ADC(2) is similar to that of sc-NEVPT2, although the former method shows a weaker active-space dependence. The ${}^{1}B_{u}$ oscillator strengths computed using the two MR-ADC methods do not change significantly with increasing active space and agree well with each other. 

\begin{table}[t!]
	\caption{Vertical excitation energies (eV) of the butadiene (\ce{C4H6}) molecule for the lowest-energy triplet state (${}^{3}B_{u}$), as well as the lowest-energy dipole-allowed (${}^{1}B_{u}$) and dipole-forbidden (${}^{1}A_{g}$) singlet states. The MR-ADC results also show the oscillator strength strengths ($f$) in parentheses. See \cref{sec:computational_details} for details of the calculations.}
	\label{tab:ee_buta}
	\setlength{\extrarowheight}{2pt}
	\setstretch{1}
	\footnotesize
	\centering
	\begin{threeparttable}
		\begin{tabular}{ccccc}
			\hline\hline
			          Method           & Active Space &             \multicolumn{3}{c}{States}              \\
			                           &              & ${}^{3}B_{u}$ &  ${}^{1}B_{u}$  & ${}^{1}A_{g}$ \\ \hline
			EOM-CCSD\tnote{\textbf{a}} &              &     $3.20$     &      $6.53$      &     $7.28$     \\ \hline
			        sc-NEVPT2          &  $(4e,4o)$   &     $3.48$     &      $5.91$      &     $6.90$     \\
			                           &  $(4e,6o)$   &     $3.45$     &      $6.40$      &     $6.86$     \\
			                           &  $(4e,8o)$   &     $3.43$     &      $6.54$      &     $6.77$     \\ \hline
			        MR-ADC(2)          &  $(4e,4o)$   &     $3.42$     & $6.13$ ($0.619$) &     $6.64$     \\
			                           &  $(4e,6o)$   &     $3.56$     & $6.46$ ($0.684$) &     $6.84$     \\
			                           &  $(4e,8o)$   &     $3.60$     & $6.52$ ($0.697$) &     $6.91$     \\ \hline
			       MR-ADC(2)-X         &  $(4e,4o)$   &     $3.26$     & $5.77$ ($0.590$) &     $6.20$     \\
			                           &  $(4e,6o)$   &     $3.50$     & $6.38$ ($0.670$) &     $6.68$     \\
			                           &  $(4e,8o)$   &     $3.54$     & $6.39$ ($0.677$) &     $6.70$     \\ \hline
			  SHCI\tnote{\textbf{b}}   &              &     $3.37$     &      $6.45$      &     $6.58$     \\ \hline\hline
		\end{tabular}
			\begin{tablenotes}
				\item[\textbf{a}] 	\footnotesize{From Ref.\@ \citenum{Copan:2018p4097}}
				\item[\textbf{b}] 	\footnotesize{From Ref.\@ \citenum{Chien:2018p2714}}
			\end{tablenotes}
	\end{threeparttable}
\end{table}

\begin{figure*}[t!]
	\subfloat[]{\label{fig:buta_bright_state}\includegraphics[width=0.45\textwidth]{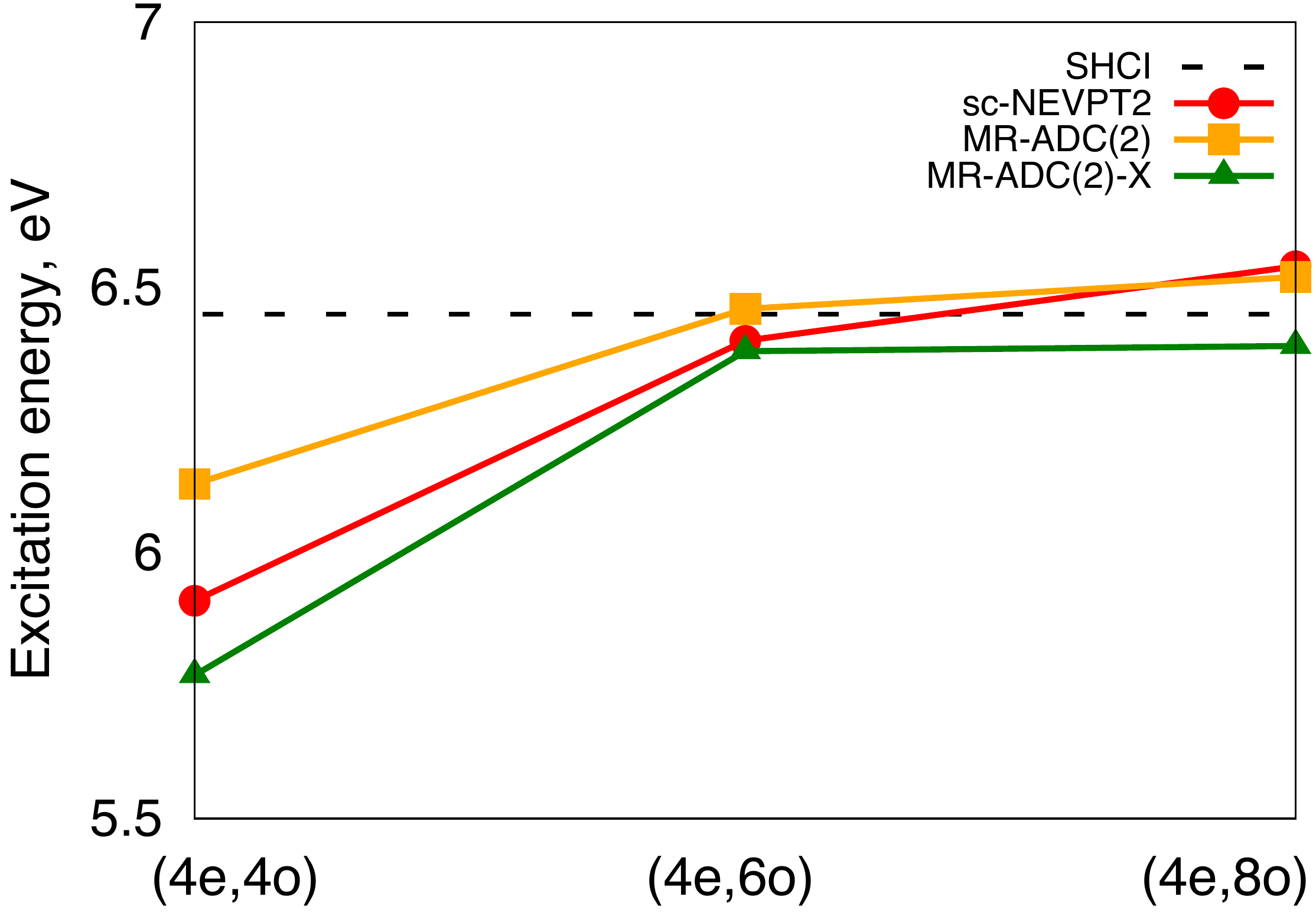}} \quad
	\subfloat[]{\label{fig:buta_dark_state}\includegraphics[width=0.45\textwidth]{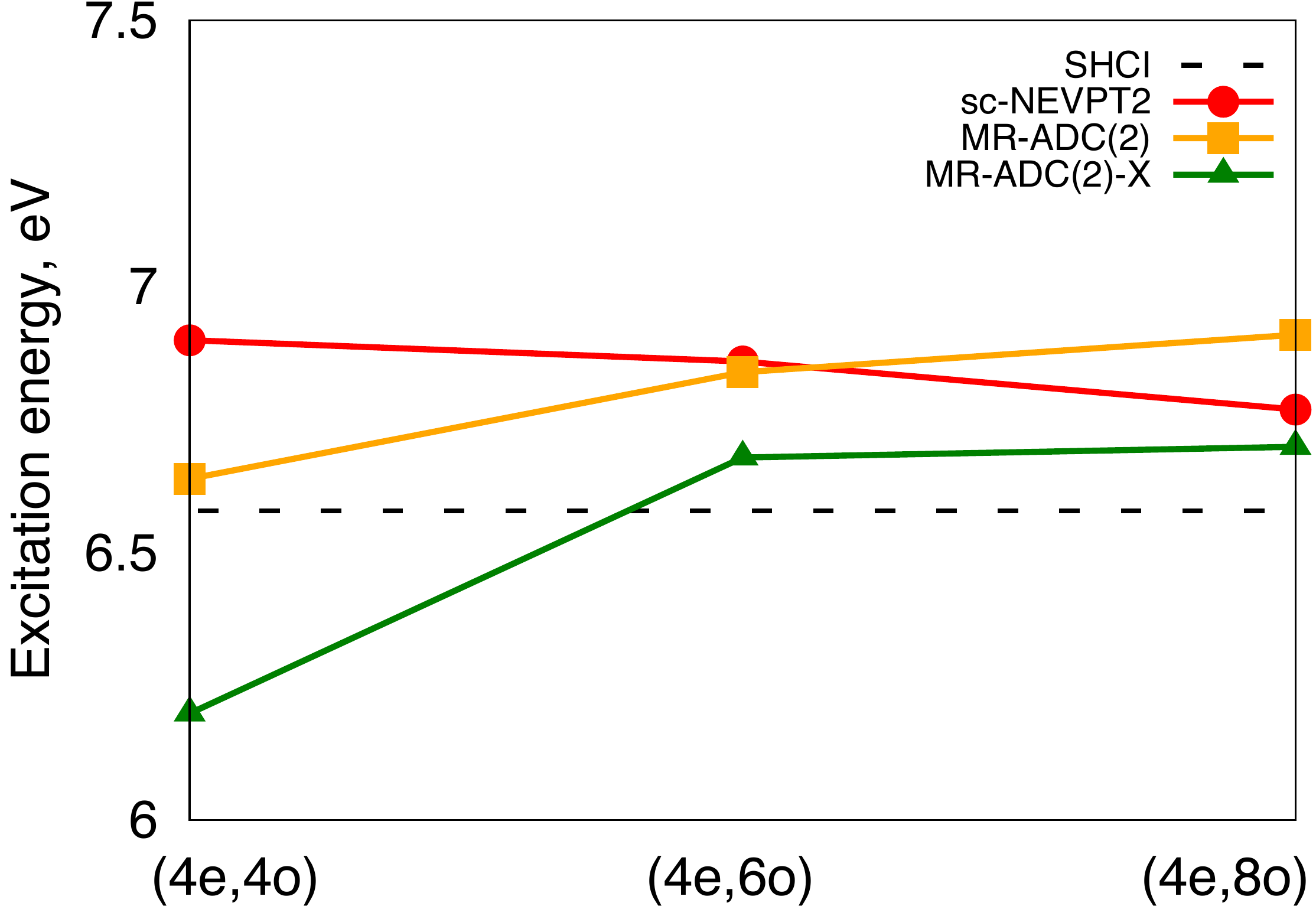}}
	\captionsetup{justification=raggedright,singlelinecheck=false}
	\caption{Vertical excitation energies for the butadiene dipole-allowed ${}^{1}B_{u}$ (plot (a)) and dipole-forbidden ${}^{1}A_{g}$ (plot (b)) electronic states computed using MR-ADC(2), MR-ADC(2)-X, and sc-NEVPT2 as a function of the number of active $\pi$-orbitals ($n$) with 4 active electrons $(4e, no)$. The reference excitation energies computed using SHCI\cite{Chien:2018p2714} are shown as dashed horizontal lines. }
	\label{fig:buta_states}
\end{figure*}	

Vertical excitation energies of butadiene are shown in \cref{tab:ee_buta}. As in our study of ethylene, we employ three active spaces for the multireference methods ranging from the full-$\pi$ $(4e, 4o)$ to the double-$\pi$ $(4e, 8o)$ space. Notably, for this system the EOM-CCSD method shows a large 0.7 eV error for the dipole-forbidden ${}^{1}A_{g}$ state, which is known to have a double-excitation character and requires a multireference treatment. The active-space dependence of the multireference methods for the ${}^{1}A_{g}$ and ${}^{1}B_{u}$ excited states is illustrated in \cref{fig:buta_states}. When combined with the smallest $(4e, 4o)$ active space, all three multireference methods outperform EOM-CCSD for the ${}^{1}A_{g}$ excitation energy with errors less than 0.4 eV. Using the largest $(4e, 8o)$ active space, the best results are shown by MR-ADC(2)-X and sc-NEVPT2 with errors of 0.12 eV and 0.19 eV, respectively. A stronger active-space dependence is observed for the dipole-allowed ${}^{1}B_{u}$ state with the sc-NEVPT2 and MR-ADC(2)-X excitation energies varying by $\sim$ 0.6 eV. Employing the $(4e, 8o)$ active space, the best results are shown by MR-ADC(2)-X that outperforms sc-NEVPT2 and EOM-CCSD with an error of 0.06 eV in the ${}^{1}B_{u}$ relative energy. sc-NEVPT2 and MR-ADC(2) overestimate the ${}^{1}B_{u}$ energy by 0.09 and 0.07 eV, respectively. The MR-ADC methods also show a significant active-space dependence in the computed ${}^{1}B_{u}$ oscillator strength that changes by $\sim$ 15 \% from $(4e, 4o)$ to $(4e, 8o)$. A weaker dependence on the active space is observed for the ${}^{3}B_{u}$ state, where the best agreement with SHCI is shown by sc-NEVPT2. Overall, the presented results demonstrate that the MR-ADC methods are competitive in accuracy with sc-NEVPT2 for all three electronic states of butadiene while capable of describing the multireference nature of the  ${}^{1}A_{g}$ state that is not captured by EOM-CCSD.

\section{Conclusions}
\label{sec:conclusions}

In this work, we presented an implementation and benchmark of the strict and extended second-order multireference algebraic diagrammatic construction theory for neutral electronic excitations (MR-ADC(2) and MR-ADC(2)-X). The MR-ADC(2) approximation incorporates all contributions to the polarization propagator up to the second order in its time-independent multireference perturbative expansion, while the MR-ADC(2)-X method includes additional third-order terms in the description of double excitations outside of the active space. We benchmarked the performance of MR-ADC(2) and MR-ADC(2)-X for the excited states in five small molecules (\ce{HF}, \ce{CO}, \ce{N2}, \ce{F2}, and \ce{H2O}) at equilibrium and stretched geometries, a carbon dimer (\ce{C2}), as well as ethylene (\ce{C2H4}) and butadiene (\ce{C4H6}). 

Our results demonstrate that the MR-ADC methods provide accurate predictions of excited-state energies for the weakly- and strongly-correlated electronic states, at geometric equilibrium or near dissociation. For the weakly-correlated systems, we find that the MR-ADC(2) and MR-ADC(2)-X methods outperform the third-order single-reference ADC approximation (ADC(3)) and are often competitive with the results from equation-of-motion coupled cluster theory (EOM-CCSD). For the multireference problems, the performance of MR-ADC(2) and MR-ADC(2)-X is similar to that of strongly-contracted N-electron valence perturbation theory (sc-NEVPT2). The MR-ADC methods have a number of added benefits, such as a straightforward and efficient calculation of excited-state properties, a direct access to excitations outside of the active space, and an ability to calculate multireference excited states without state-averaged reference wavefunctions. 

Our current work can be extended in many possible directions. One important direction is an efficient implementation of the MR-ADC(2) and MR-ADC(2)-X methods that will enable calculations of larger molecular systems with more orbitals and electrons in the active space. Another extension is to take advantage of the MR-ADC ability to calculate excitations outside of the active space, which can be used for efficient calculations of the X-ray absorption spectra of multireference systems. Finally, the methods presented in the current work can be extended to account for relativistic effects, such as spin-orbit and spin-spin coupling, which will enable calculations of magnetic systems with heavy elements. Work along these directions is ongoing in our group.

\section{Appendix: Equations for the $t_{x}^{y(2)}$ Amplitudes}
\label{sec:appendix}
As discussed in \cref{sec:theory:T_amps}, the MR-ADC(2) and MR-ADC(2)-X second-order effective Hamiltonian $\tilde{H}^{(2)}$ incorporates terms that depend on the internal second-order amplitudes $t_{x}^{y(2)} (x>y)$ that ensure Hermiticity of the effective Hamiltonian matrix $\mathbf{M}$. The $t_{x}^{y(2)} (x>y)$ amplitudes parametrize a contribution to the second-order excitation operator $T^{(2)}$ (\cref{eq:mr_adc_t_amplitudes}) that has the form
\begin{align}
\label{eq:appendix_1}
T^{(2)} 
&\Leftarrow \sum_{x>y} t_{x}^{y(2)} \c{y} \a{x} 
\end{align}
Following our previous work,\cite{Sokolov:2018p204113,Chatterjee:2019p5908,Chatterjee:2020p6343} the amplitudes $t_{x}^{y(2)} (x>y)$ are obtained by solving the projected linear equations \eqref{eq:proj_amplitude_equations} that can be written as:
\begin{align}
\label{eq:appendix_2}
\braket{\Psi_0|\c{x}\a{y}\tilde{H}^{(2)}|\Psi_0} = 0 \quad (x>y)
\end{align}
\cref{eq:appendix_2} can be simplified and converted to a tensor form:
\begin{align}
\label{eq:appendix_3}
\mathbf{K}\,\mathbf{T^{(2)}} = -\mathbf{V^{(2)}} 
\end{align}
where $\mathbf{T^{(2)}}$ contains $t_{x}^{y(2)} (x>y)$ and the elements of $\mathbf{K}$ and $\mathbf{V^{(2)}}$ are defined as follows:
\begin{align}
\label{eq:appendix_4}
K_{xy,wz} &= \braket{\Psi_0|(\c{x}\a{y} - \c{y}\a{x}) [H^{(0)}, \c{z}\a{w} - \c{w}\a{z}]|\Psi_0} \\
V_{xy}^{(2)} &= \braket{\Psi_0|(\c{x}\a{y} - \c{y}\a{x})V^{(2)}|\Psi_0} 
\end{align}
Expression for $V^{(2)}$ can be found in Eq. (38) of Ref.\@ \citenum{Chatterjee:2019p5908}.

\cref{eq:appendix_3} has the form of the Newton-Raphson equation for the second-order parameters $t_{x}^{y(2)} (x>y)$ that describe relaxation of the active-space orbitals in response to dynamical correlation from outside of the active space. It can be solved by diagonalizing the $\mathbf{K}$ matrix according to the generalized eigenvalue problem
\begin{align}
\label{eq:appendix_5}
\mathbf{K}\,\mathbf{Z} = \mathbf{S} \,\mathbf{Z}\, \boldsymbol{\epsilon}
\end{align}
and inverting the amplitude equation \eqref{eq:appendix_3} to obtain:
\begin{align}
\label{eq:appendix_6}
\mathbf{T^{(2)}} = - \mathbf{S^{-1/2}} \, \mathbf{\tilde{Z}} \, \boldsymbol{\epsilon^{-1}} \, \mathbf{\tilde{Z}^\dag} \, \mathbf{S^{-1/2}} \, \mathbf{V^{(2)}} \quad (x>y)
\end{align}
In \cref{eq:appendix_5,eq:appendix_6} the elements of $\mathbf{S}$ are defined as
\begin{align}
\label{eq:appendix_7}
S_{xy,wz} &= \braket{\Psi_0|(\c{x}\a{y} - \c{y}\a{x})(\c{z}\a{w} - \c{w}\a{z})|\Psi_0} 
\end{align}
and $\mathbf{\tilde{Z}} = \mathbf{S^{1/2}} \, \mathbf{Z}$.

\suppinfo
Benchmark of the approximation for the second-order amplitudes in \cref{eq:t2_amp_tensor_approx}. Size-intensivity tests for MR-ADC(2) and MR-ADC(2)-X. Cartesian geometries of the ethylene and butadiene molecules. Active space orbitals in the ethylene and butadiene calculations.

\acknowledgement
This work was supported by the National Science Foundation under Grant No.\@ CHE-2044648. Computations were performed at the Ohio Supercomputer Center under the projects PAS1583 and PAS1963.\cite{OhioSupercomputerCenter1987} 


\providecommand{\latin}[1]{#1}
\makeatletter
\providecommand{\doi}
  {\begingroup\let\do\@makeother\dospecials
  \catcode`\{=1 \catcode`\}=2 \doi@aux}
\providecommand{\doi@aux}[1]{\endgroup\texttt{#1}}
\makeatother
\providecommand*\mcitethebibliography{\thebibliography}
\csname @ifundefined\endcsname{endmcitethebibliography}
  {\let\endmcitethebibliography\endthebibliography}{}

\end{document}